\documentstyle[psfig]{l-aa}

\def\alwaysmath#1{\ifmmode{#1}\else{$#1$}\fi}
\def\msun{\alwaysmath{{M}_{\odot}}}

\def\etal{{et al.~}}

\def\C12C13{\alwaysmath{{}^{12}{\rm C/{}^{13}C}}}

\def\d14{\alwaysmath{\Delta V_{1.4} }}

\def\br{{\it bright}}
\def\gbr{{\it global bright}}
\def\fa{{\it faint}}
\def\gfa{{\it global faint}}
\def\hst{{\it HST~}}

\def\notbss{{\rm no}}
\def\outfld{{\rm out}}

\begin{document}

\thesaurus{07         
(08.01.1; 02.14.1; 10.05.1)} 
\title{ HST OBSERVATIONS OF BLUE STRAGGLER STARS IN THE CORE OF
 THE GLOBULAR CLUSTER M3}

\author{
F.R. Ferraro\inst{1}, B. Paltrinieri\inst{1}, F. Fusi Pecci\inst{1,2}, 
C. Cacciari\inst{1,3}, 
B. Dorman\inst{4,5}, 
R.T. Rood\inst{5}, 
R. Buonanno\inst{6},  C.E. Corsi\inst{6}, 
D. Burgarella\inst{7},  M. Laget\inst{7}
}

\institute{\inst{1} Osservatorio Astronomico di Bologna, Via Zamboni 33, 
I-40126 Bologna, ITALY\\
\inst{2} Stazione Astronomica di Cagliari, 09012 Capoterra, ITALY\\
\inst{3} Space Telescope Science Institute, Baltimore, USA \\
\inst{4} Goddard Space Flight Center, NASA Greenbelt, USA\\
\inst{5} University of Virginia, Charlottesville, USA\\
\inst{6} Osservatorio Astronomico di Roma, Roma, ITALY\\
\inst{7} Laboratoire d'Astronomie Spatiale, CNRS Marseille, FRANCE}

\offprints{ F.R. Ferraro, e-mail ferraro@astbo3.bo.astro.it }

\date{}

\maketitle
\markboth{Ferraro \etal}{Blue Stragglers in M3}

\begin{abstract}

The core of the Galactic Globular Cluster M3 (NGC 5272) has been observed with
the WFPC2 through the filters $F255W$, $F336W$, $F555W$, and $F814W$.
Using these observations along with a thorough reanalysis of earlier
catalogs, we have produced a catalog of blue straggler stars (BSS)
spanning the cluster. Earlier studies and the fainter part of our sample
suffer severe selection biases. Our analysis is based on a more reliable
{\it bright global sample} of 122 BSS. We confirm earlier suggestions that
the radial BSS distribution in M3 is bimodal. It is strongly peaked in the
center, has a clear dip 100--200\arcsec\ from the center, and rises again
at larger radii. The observed distribution agrees with the dynamical model
of Sigurdsson \etal\ (1994) which takes into account both star collisions
and merging of primordial binaries for the origin of BSS. The observed
luminosity functions of BSS in the inner and outer parts of the cluster
are different. Interpreting these using the models of Bailyn \&
Pinsonneault (1995), we suggest that the BSS in the inner cluster are
formed by stellar collisions and those in the outer cluster from merging
primordial binaries.

\end{abstract}

%

\section{Introduction}

With the advent of high-resolution, highly accurate imaging facilities,
there has been a growing interest in the systematic search and study of
Blue Straggler Stars (BSS) in Galactic Globular Clusters (GGCs). BSS were
first detected in the GGC M3 by Sandage (1953) and by 1995 there were more
than 700 BSS candidates identified in about 35 GGCs (see the review of
Bailyn 1995 and the catalogs of Ferraro \etal\ 1993, 1995, Sarajedini
1993).  The BSS sample is increasing rapidly as
more GGC cores are adequately surveyed with the {\it Hubble Space Telescope}
(\hst).

Even with the dramatic increase in the known BSS population, the most
interesting cluster population of BSS remains that in M3. This 
is the only bright cluster where the radial distribution of BSS has
been studied extensively. In particular, Ferraro \etal\ (1993, 
hereafter F93) found a
bimodal radial distribution of the BSS candidates. This could arise if
there were different BSS formation mechanisms at work in
regions of different stellar density. For example, the external areas would
contain mostly BSS that were formed in primordial binary mergers, whereas
the core would contain mostly BSS formed via collisions (see Bailyn \&
Pinsonneault 1995, hereafter BP95, and Bailyn 1995, for a general 
discussion and references).
Alternatively, one could imagine that, though formed via the same
mechanism, the BSS have been subject to radial mass
segregation or suffered different disruption/survival histories.

This situation is apparently unique to M3; no other cluster has been found
to show this bimodal distribution of BSS. In addition M3 serves as an
archetype for several astrophysical problems---e.g., its ``horizontal''
horizontal branch gave the name to the sequence. It has been well studied
both photometrically and spectroscopically. Because of these factors
we have selected M3 as a primary target for a long term project
investigating the global populations of globular clusters, in spite of
the fact that it is not the nearest cluster in the sky.

 During the last decade we have secured in M3 one of the {\it widest and
most complete} samples of stars ever observed in a globular cluster using
different techniques (photographic plates:  Buonanno \etal\ 1986, 1994
[PH94], and CCD-arrays:  Ferraro \etal\ 1997 [CCD97]), covering most of the
cluster area (from 20\arcsec\ up to 7\arcmin\ from the cluster center).
To achieve the ultimate in sample size and completeness and to check for
radial variations in the cluster population, we obtained high-resolution
\hst\ observations during Cycle 4 (GO 5496 P.I. F. Fusi Pecci) aimed at
completing the survey in the inner regions.  The project as a whole aimed
at testing the accuracy and validity of the assumptions and physical
inputs that are the foundation of the current stellar evolution theoretical
models. 

This paper is the first of a series reporting on the results obtained from
the global Colour-Magnitude Diagrams (CMDs) and Luminosity Functions (LFs).
Here we present results on BSS candidates in the cluster core and discuss the
distribution of BSS across the cluster.

\section{Observations, data reduction}

A more detailed descriptions of observations and reductions
will be given elsewhere (Fusi Pecci \etal\ 1997). We outline here the
important points.

The \hst\ frames were obtained on 1995, April 25 (GO 5496) with the WFPC2
coupled with the filters $F255W$ ($m_{255}$), $F336W$ ($U$), $F555W$
($V$), and $F814W$ ($I$).  The PC was roughly centered on the cluster
center, while the WF cameras partially overlapped outer regions previously
observed from the ground (see Figure 1). While complete reductions have
already been carried out for most available frames, here we make use only
of the $F255W$, $F336W$ and $F555W$ observations for all four chips (PC, WF2,
WF3, WF4) and $F814W$ data for the PC.  The list of the frames and the
exposure times per frame are reported in Table 1.

Each WFPC2 frame was processed through the standard HST-WFPC2 pipeline 
for bias subtraction, dark correction, and flat-fielding.
Then, all the long exposure images were registered using  MIDAS,
obtaining a median frame where the cosmic rays have been removed.
We used the median frame  in order to search for all the individual
star components present in each field 
following  the standard procedure implemented in ROMAFOT (Buonanno \etal\ 
1993), a package specifically developed for high precision photometry in 
crowded fields.
The PSF fitting photometry was performed on each individual frame 
separately, and an averaged instrumental magnitude was computed
for each star.
The instrumental magnitudes ($m_{\rm instr}$) were
finally transformed to the Johnson system 
(for $V,I,U$, respectively) using formula 8 and Table 7 from
Holtzmann et al. (1995), and in the STMAG system using Table 9 for the
$F255W$ filter.

Figure 2 shows the CMD ($m_{255},~ m_{255}-U$) for more than 18,000 stars
measured in the observed area.

\begin{table}
\begin{center}
\label{t:obslog}
\caption[]{ List of the used frames and exposure times per frame. }
\begin{tabular}{cccc}
\hline\hline
\\
Filter  & No. Expos. & Exp.Time (sec) & Date \\
\\
\hline
\\
F255W  &  4 & 300 & April 25,1995\\
F336W  &  4 & 800 & April 25,1995\\
F336W  &  2 &  70 & April 25,1995\\
F555W  &  4 & 100 & April 25,1995\\
F555W  &  2 &   3 & April 25,1995\\
F814W  &  4 & 140 & April 25,1995\\
F814W  &  2 &   3 & April 25,1995\\
\\
\hline
\end{tabular}
\end{center}
\end{table}

\section{The BSS sample from our \hst\ observations}

As already shown in other UV studies of GGCs (Dorman, Rood, O'Connell 1995
and references therein), the main characteristic of the UV CMD is that at
these wavelengths the cluster light is dominated by hot stars,
specifically the blue HB stars. The BSS are the next hottest
sub-population in the cluster, and are thus easily distinguished
from the cooler stars of the turnoff and SGB.
 The location and morphology of the main branches are very
different from the optical CMD. The RGB is very faint, and the choice
of ($m_{255}$,$m_{255}-U$) makes the HB appear diagonal.
As can be seen in Figure 2, there is a well-defined, narrow
 branch extending upward to the left from the giant
branch [note however that the choice of the colour frame makes
the HB diagonal: the HB is different again in the (U,$m_{255}-U$) plane, 
almost horizontal]. 
This branch corresponds to a large part of the HB, excluding the
hottest section, which bends downward in $m_{255}$ because of the
increasing bolometric correction.  The overall morphology of the CMD will
be discussed in detail in a future paper (Fusi Pecci \etal\ 1997), where
we will present also a comparison with theoretical models. Here, we only
discuss the BSS.

Defining a sample of candidate BSS via photometric techniques 
is somewhat arbitrary, mostly because of the difficult separation 
between the faint BSS population and the ``normal'' MS stars at the TO. 
We anticipate that, coupled with incompleteness, this problem is
one of the main reasons why samples obtained by different observers 
match so poorly.  

To minimize this problem, we have selected the BSS candidates in 
the UV CMD ($m_{255},~m_{255}-U$). As can easily be seen in Figure 2,
in this plane the BSS sequence is quite distinct, spanning 
$\sim3\,$mag in $m_{255}$. To illustrate our selection criteria better,
Figure 3 shows a zoomed CMD of the BSS region where the selected candidates 
are indicated. To reduce the impact of the selection bias on
the comparison with previous searches and on the following discussion,
we have divided the total BSS sample into two sub-samples:

\begin{enumerate}

\item \br\ BSS:  with $m_{255}<19.0$

\item \fa\ BSS:  with $19.0<m_{255}<19.4$

\end{enumerate}

\noindent
Since no gap is evident between the BSS sequence and the ``normal'' 
stars located in the TO region, we set the limiting magnitude to be
$m_{255}=19.4$ ($\sim 5 \sigma$ brighter than the magnitude of the MSTO).
The separation between the \br\ and the \fa\ samples 
has been taken to be $m_{255}=19.0$ in order to be consistent with the 
limiting magnitude adopted for the two corresponding samples in F93 
(see below).

There are several intriguing objects which are clearly located outside
the bulk of the population of the main branches and far in colour from
the BSS sequence as well. As can be seen from Figure 2, we have not
included most of these stars in our sample. They are sometimes called
{\it yellow stragglers} (McClure \etal\ 1985) as they are intermediate
between the blue stragglers and the red subgiants. One should note
that they could, and in our view probably do, ``contain'' a BSS candidate
as one component of an {\it optical} blend with a subgiant (see
Ferraro \etal\ 1992a,b).  Even though we are using high resolution \hst\
frames some optical blends are still expected because of the high
degree of crowding in the very inner regions of M3.

Some similar optical blends are present in almost
all CMDs which have been used in the search for BSS candidates, 
and they have been usually discarded. Therefore, it
is conceivable that several BSS are still ``hidden'' in these
optical binaries. Of course, one cannot even exclude that they
might be {\it evolved} BSS (see Ferraro et al 1992a,b for a discussion),
but at this stage we adopted the usual definition of BSS and have
excluded them from our reported sample.

With the above criteria and caveats, the total number of BSS in our 
\hst\ field is 171. Of these, 72 belong to the \br\  sample 
and 99 to the \fa.
The two subsamples are plotted with different symbols in Figure 3
(big dots: \br; big asterisks: \fa, respectively).
Tables 2 \& 3 list the BSS candidates: the first column is our identification 
number, then in columns 2, 3, 4, 5 we report the $V,~I,~U,~m_{255}$ magnitudes,
respectively, and in column 6, 7, 8 the coordinates ($X, Y$) and
the distance ($r$) from the adopted cluster 
center in arcsec, respectively\footnote{Complete tables 2,3,4 are
avaliable on electronic form at the Center de Donn\`ees de Strasbourg (CDS)
and can be obtained by anonymous ftp copy.}. 

The coordinate system adopted here is the same as used in our previous
papers on M3 (F93,PH94,CCD97). A further revision of the whole problem,
and a discussion on the choice of an {\it absolute} reference system is
postponed to a forthcoming paper currently in preparation (Laget \etal\ 1997).

\section{Recent searches for BSS candidates in M3}

Since a revision and discussion of available ground-based surveys 
was presented in our previous papers (F93, CCD97), 
we focus here just on the latest surveys with the 
goal of achieving a {\it revised global} sample of BSS in this cluster.

\subsection{{Ferraro \etal\ 1993 - (F93)} }

Based on independent CCD observations at the CFHT and collection and revision 
of previous data, F93 presented an extensive study of the 
BSS in M3. 

In that study 70 new BSS candidates were identified at $r<200\arcsec$.
Coupled with the 76 candidates known from 
the previous photographic samples (Sandage 1953, Sandage and Katem 1982,
Ables \etal\ 1982, PH94) this yielded the most populous data-set of known BSS
in a single GGC. On the basis of this catalog we studied the global BSS 
radial distribution setting two preliminary constraints:

\begin{enumerate}

\item In order to have an homogeneous sample we did not consider
candidates in the central region at $r<20\arcsec$ since the sample
in that region is likely to be very incomplete in any ground-based
photometry, due the high crowding conditions. Moreover, because of
inhomogeneity of the surveys, we also excluded the 14 BSS candidates
identified at very large distances ($r>360\arcsec$) (Sandage 1953,
Ables \etal\ 1982, see F93 Table 1).  Thus, the
radial extent of the adopted sample was limited to the range
$20\arcsec<r<360\arcsec$.

\item Since the photographic survey was complete to $B=18.6$, we also
limited the analysis only to the bright BSS sample ($B<18.6$).  This
choice is surely quite restrictive. On the other hand, as already
stressed, the selection of the faint BSS is difficult and quite
subjective and may thus introduce significant bias in the analysis.

\end{enumerate}

The main result presented by F93 is reported in their Figure 9: {\it the
relative frequency of the BSS in M3 displays a sharp bimodal radial
distribution, with a distinct dip in the region with
$100\arcsec<r<200\arcsec$}. However, it was evident that the BSS
distribution in the inner regions required much better observations and
F93 noted the desirability of additional independent observations.

\subsection {{Bolte \etal\ 1993 - (BHS)} }

Almost simultaneously with F93, Bolte \etal\ (1993, hereafter 
BHS) presented the
results of a BSS search in M3 carried out using the high-resolution CCD
camera at the same telescope (CFHT), but in slightly better seeing
conditions.  The area covered by BHS is shown in Figure 1 ($\sim
2.2\arcmin \times 2.2\arcmin$), and overlaps both the
previous studies and also our present
\hst\ survey.

In this very central region, BHS reported the identification of 46 BSS,
and defined a {\it specific frequency} $F_{\rm BSS}$ as the ratio of the
number of BSS with respect to those of HB and RGB stars (with
$V<V_{\rm HB}+2$):  

\begin{displaymath}
F_{\rm BSS} = {{N_{\rm BSS}} \over {N(V<V_{\rm HB}+2})}
\end{displaymath}

\noindent In
particular, for the central regions of M3 they obtained:  

\begin{displaymath}
F_{\rm BSS}^{\rm in} = {{46}\over{1382}}=0.033\pm0.005 
\end{displaymath}

\noindent Using the data presented by Paez \etal\ (1990) on a small
external area located at $\sim375\arcsec$\ from the cluster center
($r\sim15r_c$, where $r_c$ is the core radius, i.e., outside the region
considered in F93), they obtained:  
\begin{displaymath}
 F_{\rm BSS}^{\rm out} = {{9}\over{94}}=0.09\pm0.03 
\end{displaymath} 
\noindent
and concluded that the specific frequency of BSS within $5r_c$ of the 
center of M3 is a factor $\sim2.8$ smaller than that seen in the outer regions.
If confirmed, this result would be quite surprising and interesting as
{\it none} of the GGCs observed so far has shown a larger frequency 
of BSS in the outer regions than in the inner ones.

As noted in F93, this result is perhaps somewhat over-interpreted as there
was the obvious possibility that their sample could be substantially
incomplete in the innermost region of the cluster ($r< r_c$).  In this
respect, F93 found 14 BSS with
$r<25\arcsec$, a number comparable in size to the 19 found by BHS over
the same radial interval (see their table 1). However, there were only a
few coincidences between the two samples.

CCD97 have further discussed the problem and carried out
a detailed comparison between the F93 and  BHS samples for $r>20\arcsec$,
excluding the innermost region since it required a much better
spatial resolution. It has been clear that a reliable
revised sample could only be achieved after the analysis of
\hst\ data. 

\noindent
\subsection{{Guhathakurta \etal\ 1994 - GYBS }}

The first direct confirmation of the large incompleteness of the BHS data
came from the first \hst\ observations (with WFPC1) of the
central regions of M3, before the refurbishment mission. Guhathakurta
\etal\ (1994, hereafter GYBS) presented $V$ ($F555W$) and $I$ ($F814W$)
photometry of a $\sim 65\arcsec \times 65\arcsec$ region centered 
$\sim20\arcsec$ E of the commonly adopted cluster center.

They also retrieved $U$ ($F336W$) images from \hst\ WF/PC Instrument
Definition Team archive which had a small region ($\sim 25\arcsec\times
25\arcsec$) in common with their $V$, $I$ observations (see their Figure
2).  Since the ($V,~U-I$) plane is more efficient in 
searching for BSS candidates than the traditional ($V,~V-I$) CMD, they
restricted their search to only this small overlapping region (namely
$r<20\arcsec$). The approximate area of this search is plotted in our Figure
1 together with those of the other surveys. In this area GYBS
found 28 BSS candidates. 

To better understand the difficulties undermining the detection of
BSS in cluster cores, it is quite interesting to note that only half of
the 28 GYBS candidates fall in the general BSS area in the BHS CMD,
and only 30$\%$ actually fall inside the BHS adopted BSS-box (see Figure
10 in GYBS). Moreover, 10
out of 28 GYBS BSS were completely missing in the BHS survey.

As noted by GYBS (see their Figure 9 and 10), from the comparison above,
the degree of completeness achieved by BHS in that
very central area was less than 25$\%$ in the range
$V=17.5$--18.5. GYBS also computed the specific frequency of BSS in the
central region, yielding:

\begin{displaymath}
F_{\rm BSS}(r<20\arcsec)=28/297=0.094 \pm 0.019  
\end{displaymath}
\noindent
which is $\sim 3$ times the value found by BHS, and quite comparable 
to the BSS frequency they estimated in the outskirts of the cluster.

\subsection{{Burgarella \etal\ 1995 - (BPQ) }}

Burgarella \etal\ (1995, hereafter BPQ), presented UV ($F220W$, $F346W$) high
resolution HST/FOC---pre-COSTAR---observations of the core of M3 secured
before the repair mission. In the small FOC/96 field of view
($\sim11\arcsec\times 11\arcsec$), they identified 12 BSS and computed
the specific frequency which turned out to be:

\begin{displaymath}
  F_{BSS}(r<5\arcsec) = 12/42 = 0.29 \pm 0.09   
\end{displaymath}

\noindent This figure is three times the value found by GYBS and
$\sim10$ times that found by BHS. It was thus clear that increasing the
spatial resolution of the available observations increases (greatly) the
total number of detected BSS candidates.

One has to note that only 5 out of the 12 BSS candidates 
found by BPQ were also identified as BSS by GYBS, confirming yet again the
difficulties in establishing homogeneous the selection criteria. In particular,
most of the BPQ BSS candidates are actually close to the TO region 
(see Figure 3 in BPQ).

\section{The BSS specific frequency in the inner regions ($r<20\arcsec$)
and comparisons with results from previous surveys}

Before computing our independent determination of the BSS specific
frequency and carrying out any comparison with 
the available data-sets, it is important to 
stress an additional factor which has to be always kept in mind 
in the analysis and discussion.

\subsection{A basic preliminary caveat}

>From the brief review in Sect. 4 it is evident that, besides the obvious
observational problems, the definition of a {\it reliable and complete}
sample of BSS candidates in a GGC is undermined by the intrinsic
uncertainties in the definition of what a {\it blue straggler} actually is.

Drawing a BSS ``box'' is highly arbitrary.  This problem is clearly
evident from the inspection of Figure 9 and 10 in GYBS, and Figure 19a,b
in CCD97. In particular, the most critical assumptions
concern the separation of the BSS from the ``normal'' TO and subgiant
stars. Furthermore, the BSS selection is strongly dependent on the bands
employed in the CMD. In this respect, there is no doubt that
long colour baselines (GYBS) and UV filters (BPQ) are 
preferable.

For this reason we will use our new \hst\ sample to compute
the BSS specific frequency  
in the very central regions. Then, using what we have learned, 
we will construct  a {\it revised global} catalog
of BSS candidates in M3 over the complete radial range (see Sect. 6).

\subsection{{Bolte \etal\ 1993} }

As described in the previous section, BHS identified 
46 BSS candidates in the very central area. The region covered by our 
\hst\ central field F1 (see Figure 1) overlaps only 75$\%$ of the total
central zone surveyed by BHS, who have kindly made available to us 
their data in machine-readable form.

In the region in common, the number of objects listed by BHS is
$ N_{\rm BSS}^{\rm BHS}= 40$ (6 of their candidates located outside 
our field), while we have $ N_{\rm BSS}^{\br}=67 $ (and only 4 \br\ BSS
candidates are outside the BHS field). Only 30 
of all these stars are considered to be BSS in both samples.
There are 4 more BHS candidates included in our \fa\ sample, and 6 of
the BHS BSS have
been found to be RGB or SGB stars in our photometry.

In summary, our BSS central population is $\sim 1.8$ larger
than the corresponding one detected by BHS. If we considered 
as ``truly reliable'' BSS only the candidates in common between the two
searches, the ratio would rise up to 2.2. From this we estimate that 
the degree of completeness of BHS down to $m_{255}=19$ is 45--50\%. 

Turning to the specific frequency of BSS stars as defined by BHS,
on our whole \hst-F1 we have the following results:
\begin{displaymath}
F_{\rm BSS}= 72/1140 = 0.06\pm0.01
\end{displaymath}

\noindent
while in the region in common with BHS we obtain 
\begin{displaymath}
F_{\rm BSS}=67/1099=0.06\pm0.01
\end{displaymath}

\noindent
This value is almost twice the value obtained by BHS. The difference
arises from two (partially compensating) effects. 
The first, leading to increase  $F_{\rm BSS}$, is the larger
number of BSS we detected; the second, leading to decrease $F_{\rm BSS}$, comes
from the higher completeness achieved in the detection of
the ``normal'' reference stars.

Most of the high resolution studies in M3 consider
the specific frequency within the innermost region with $r<20\arcsec$,
where we 
get: 
\begin{displaymath}
 F_{\rm BSS}(r<20\arcsec) = 32/290=0.11 \pm 0.02  
\end{displaymath}

\noindent
And from
BHS sample, we find:
\begin{displaymath}
 F_{\rm BSS}^{\rm BHS}(r<20\arcsec) = 21/315=0.07 \pm 0.02
\end{displaymath}

\noindent a figure which is surprisingly twice the value they reported on
the whole area.  Note that using {\it their} sample and {\it their} 
assumptions on the
cluster center (485,~689 in their coordinate system), we found 20 BSS with
$r<20\arcsec$ and 24 BSS with $r<25\arcsec~(= 1r_c)$, which is quite
different from that listed in their Table 1. 

This evidence indicates that
in the BHS sample $F_{\rm BSS}$ is much higher in the very innermost
region than averaged over the whole sampled area. In particular, we found
$ F_{\rm BSS}^{\rm BHS}(r>20\arcsec)= 26/1071=0.024\pm 0.005$, which is a
factor of 3 smaller than that obtained in the inner region. 

We can thus conclude that, contrary to their own conclusions,
 the BHS data already showed a 
high BSS specific frequency in the very inner regions, even though their
achieved completeness was still quite low.

\subsection{{Guhathakurta \etal\ 1994} }

The ratio obtained in the innermost region of M3 ($r<20\arcsec$) from our
\hst\ sample is compatible with the one found by GYBS.  The number of BSS
detected in the common area is almost equal at similar magnitude levels
(28 by GYBS, 32 in our \br\ sample and 44 in the \fa\ one).  However, only
24 of these candidates are in common between the two independent
selections (19 in the \br\ sample and 5 in the \fa\ one, respectively).

\subsection{{Burgarella \etal\ 1995} }

In the small region covered by their FOC/96 observations, BPQ
detected 12 BSS. In the same zone, we identified 17 candidates 
(8 in the \br\ sample and 9 in the \fa\ one, respectively)
but only 8 
(4 in the \br\ sample and 4 in the \fa, respectively)
are in common.
We also measured many more ``normal'' stars so that, though increasing
the number of BSS, we actually get a lower $ F_{\rm BSS}$ than found by BPQ
(if only {\it bright} BSS are counted):
\begin{displaymath}
 F_{\rm BSS} = 8/52 = 0.15 \pm 0.08 
\end{displaymath}
\noindent
which is still much higher than that obtained by averaging over the
total central area. 
The BPQ selection of BSS candidates has been performed using an UV
CMD down to $m_{220}<19.0$ which is almost coincident with the
magnitude ($m_{255}<19.0$) we adopted to separate the \br\ and \fa\
samples. Despite this the final lists are significantly different.
This is partially due to the differences in the response of the used
cameras (FOC/96 + $F220W$ and WFPC2 + $F255W$), but it certainly
confirms the importance of the photometry and the adopted selection
criteria, even starting from quite comparable observational set-ups.

In fact, if we consider
our global sample in the region covered by their field we would get:
\begin{displaymath}
 F_{\rm BSS} = 17/52 = 0.33 \pm 0.08 
\end{displaymath}
which is fully consistent with the value (0.29) they obtained.

As first noted by BPQ, these values indicate that the BSS 
specific frequency significantly increases toward the center of M3.
Still the small numbers of stars and the residual incompleteness 
(for both BSS and reference stars) yield an uncertainty large enough 
to make these values compatible with the $F_{\rm BSS}$ obtained in the 
whole region with $r<20\arcsec$.

\section{The adopted global BSS sample in M3}

>From the discussion above it is clear that the possible biases
affecting any sample of BSS candidates make it quite difficult to
produce a {\it reliable} global catalog of the BSS spanning the
complete radial range from the very center out to the far outskirts of
M3.

We have therefore decided to collect all the candidates proposed so
far by any search and find all the coincidences among the various
lists.  At this point we adopt general criteria to select a {\it
fiducial} list of BSS candidates. It is not trivial to establish
criteria which would give {\it safe} BSS identifications.

One cannot simply select those stars labeled as BSS in at least two
independent surveys, because the different surveys have different
resolution and sample different regions of the cluster with different
crowding conditions. UV high resolution \hst\ data in the central
regions yield much more efficient and reliable detections.

In order to make appropriate choices taking into account the above
caveats, we adopted different selection criteria in three different zones:

\begin{enumerate}

 \item The inner region ($r<20\arcsec$), with 4 different surveys
(three from \hst, namely, GYBS, BPQ and this study, and one from
the ground, BHS). In this region we have considered as {\it reliable}
BSS candidates only the objects with at least 2 (out of 4 possible)
independent identifications.

\item  An intermediate region, where HST observations
(from this survey) and two ground-based surveys (BHS and F93) are available. 
In this zone we exclude from the {\it reliable} BSS sample any 
candidate from ground-based observations which is not confirmed by the HST 
independent search.

\item An additional (quite small) region, with 2 ground based surveys
(BHS and F93), but with no HST observations for verification.  The
spatial resolution of these studies is sufficiently high, considering
the much lower degree of crowding at these radial distances from
the center, so we accepted all the BSS candidates identified in this
region by both surveys.

\end{enumerate}

Table 4 reports the total list of the 263 selected BSS candidates,
referred to the same coordinate system (see PH94) 
and {\it roughly} to the same photometric system for the $V$ band,
eventually applying the magnitude shift (discussed in CCD97) to GYBS and 
BHS original magnitudes and colours. 

The various columns are: 
(1) the new identification number (in order of increasing distance from 
the cluster center); 
(2) the identification numbers (when available) in: this paper 
(HST, Table 2, 3), BPQ, GYBS, BHS, F93, Paez \etal\ (1990), 
Ables \etal\ (1982), Sandage and Katem (1982), and Sandage (1953), 
respectively; 
(3) the (apparent) V magnitude adopted to study the BSS luminosity function 
(note that original magnitudes and colours can be found in each quoted study);
(4)-(5) the $X, ~Y$ coordinates in arcsec; 
(6) the distance ($r$, in arcsec) from the adopted cluster center;  
(7)-(10) the identification labels: 
numbers indicates the identification numbers in each study in which the BSS
has been detected;
the label ``\notbss'' means the BSS candidate is located in the field of view 
of that survey but it was not identified as a BSS; ``\outfld'' indicates the 
BSS is located outside the field of view. 

Since there is a large difference in the quality of the sample with
decreasing the BSS luminosities, and since we eventually aim at
studying the radial distribution of the bright BSS (to avoid strong
MS contamination and poor photometry) as we did in our previous
study (F93), we have divided also the global sample into two
sub-groups, according to the BSS luminosity. In most of the following
discussion we will concentrate on the {\it global bright}
sample (whose 122 members are flagged in Table 4).

The \hst\ observations of the core of M3 have allowed us to 
complete the BSS survey over the entire cluster. Since
the field covered by the PC is almost exactly the same as that of
the data set presented by F93 and CCD97, the main problem 
we have here is to match the various
photometric systems so that the luminosity cuts in the
original sub-samples are equivalent.  This
procedure may have a quite significant impact on the discussion of
the radial distribution since, usually, different detectors and photometric
systems have been used in different cluster zones.

Using the BSS having all the necessary colours, we have estimated that the
limiting magnitude used in F93 ($B=18.6$) to separate the bright and faint
BSS samples corresponds to the magnitude $V \sim 18.3$, and to 
$m_{255}=19.0$ used in Sect. 3 to separate the two corresponding 
\hst\ samples. 
To do this we traced a mean ridge line for the BSS locus in the
various CMDs and read the luminosity correspondence at fixed colour.  Since
the BSS in clusters typically display a wide spread in colour the use of
the ridge line would be not strictly justified. However, we believe that 
adopting the limits $m_{255}=19.0$, $B=18.6$, and $V\sim 18.3$ provides 
a homogeneous cut with a precision of $\pm 0.10\,$mag or better.

\subsection{The adopted \gbr\ BSS sample}

The 122 BSS brighter than the above magnitude limits (included in Table 4) 
over the total radial range form our {\it global bright BSS sample}. 
In addition, based on the criteria
adopted in F93, we have selected samples of reference ``normal'' stars
spanning a similar magnitude interval. These are essentially subgiant
branch stars, whose samples should be as complete as the BSS population
detected in the same areas, with the same search techniques.

Since the positions of the RGB reference stars in the external region
($r>360\arcsec$) are not available from Sandage (1953), we limited the
following analysis to the area within 0--360\arcsec.

Over this area we have measured $V$ magnitudes (though in different
photometric systems from ground-based or \hst\ observations) and one or
more colours ($m_{255},~B$ and/or $I$) for almost all these stars.  In
Figure 4 we plot CMDs in the appropriate colours in different radial
regions for both the BSS ({\it filled triangles}) and reference stars 
({\it dots}).  {\it Panel (a)} shows the $(V,~m_{255}-V)$ CMD for the
region covered by \hst -- the two lines indicate the limiting magnitudes 
for the RGB ($B<18.6$) and the BSS ($m_{255}<19.0$) samples.  
{\it Panel (b)}
shows the $(V,~V-I)$ CMD for the region with $r<210\arcsec$ and outside
the \hst\ field. {\it Panel (c)} shows the $(V,~B-V)$ CMD for the outer
regions (out to r$<360\arcsec$). In {\it panel (c)}, 7 bright BSS from
Sandage (1953) lying at $r>360\arcsec$ have been plotted as open squares, for
completeness. Their original colours have been shifted ($\Delta V=0.077$
and $\Delta (B-V)=0.15$) to fit to the BSS sequence.

In Figure 4(a), two bright BSS candidates (namely $No.$ 6430 and 35060) have 
$m_{255}-V>1.5$; these stars have been found to be strongly contaminated 
in the $V$ band by nearby ($<0.2\arcsec$) red bright stars.

\subsection{The adopted \gfa\ BSS sample}

As already stressed, the reliability of the \gfa\ sample is low due to the
strong selection bias. One should use it with particular care. 
The present
sample gives 141 faint BSS candidates.  Their radial distributions in the
four annuli considered to describe the \gbr\ sample would be 43, 59, 34, 5
at $r<20\arcsec$, $20\arcsec<r<210\arcsec$, $210\arcsec<r<360\arcsec$
and $r>360\arcsec$, respectively. However, since the quality of BSS
searching, photometry, and selection is highly variable with distance from
the cluster center, we preferred to make no further use of this sample.

\section{The BSS radial distribution} 

As pointed out by many authors and recently summarized by Bailyn (1995),
most BSS in GGCs are found to be centrally concentrated with respect to
``normal'' stars. Since the central relaxation time in these systems
is much smaller than the cluster age, this result is generally ascribed
to dynamical mass segregation and interpreted as an evidence that BSS are 
more massive than the comparison stars.
At variance with this simple scenario, in F93 we found that the radial 
distribution of BSS in M3 is clearly bimodal (see F93, Figure 9).

In this section we discuss first the radial distribution of BSS in 
the inner regions observed with \hst, and then extend the analysis
to the \gbr\ sample spanning the whole radial range. To this aim
we have applied the procedure already discussed in our previous 
papers (see F93, CCD97). 

\subsection{The BSS radial distribution: the \hst central sample}

The radial distribution of the 72 \br\ BSS candidates (with
$m_{255}<19.0$) listed in Table 2 has been compared to that of a sample of
RGB stars assumed as ``reference'' population.  The BSS have been selected
using the $m_{255}$ magnitudes in the UV-CMD, and the RGB stars are much
fainter in this band. To compensate for this we checked the reference
sample in the ($V,~V-I$) CMD, eventually choosing RGB stars brighter than
$V\sim17$. This procedure reduces the (small) fluctuations
introduced by the poorer photometry for the very red stars in the UV bands.

The cumulative radial distributions for the 72 \br\ BSS and the 567 RGB
stars are plotted in Figure 5 as a function of their projected distance
from the cluster center.  It is evident from the plot that the BSS (solid
line) are more centrally concentrated than RGB stars (dotted line). In
fact, $\sim 50\%$ of the \br\ BSS are inside $r<24\arcsec$ ($\sim 1r_c$),
while only $\sim 33\%$ RGB stars are located within this distance.

A Kolmogorov-Smirnov test has been applied to the two
distributions to check the statistical significance of the detected
difference. The test yields a probability of $\sim 99.5\%$ ($\sim3\sigma$
level of confidence) that the \br\ BSS population in the central region 
of M3 has a different radial distribution than the selected RGB stars.
The level of confidence grows to $\sim3.5\sigma$ if one considers the 
whole sample of 171 (\br+\fa) BSS listed in Tables 2 and 3.

In conclusion, {\it there is a significant evidence that the \br\ 
BSS candidates are more centrally concentrated than the RGB
stars spanning the same magnitude interval.}

\subsection{The BSS radial distribution: the global sample}

To study the radial distribution of the \gbr\ sample, we have 
compared its cumulative radial distribution with that of the
reference stars. The two
groups of stars consist of 114 BSS 
(out of 122) with $r<360\arcsec$ and 1581 RGB stars, and
their cumulative radial distributions are reported in Figure 6.
As it is evident from the plot and already shown in our previous
study (F93, Figure 7), there is a clearly bimodal trend, with
the BSS (solid line) more centrally concentrated than RGB stars 
(dotted line) in the central regions (out to r$<100\arcsec$), while
the opposite occurs in the outer regions.
Figure 7 reports the cumulative distributions for the data broken into
two subsets, separated at $r=150\arcsec$. 

Kolmogorov-Smirnov tests  applied to RGB and BSS distributions yield
the following results for the significance of the difference:

\begin{enumerate}

 \item global sample ($0\arcsec<r<360\arcsec$): $\sim 99.96\%$
($\sim3.5\sigma$)

 \item inner sample ($0\arcsec<r<100\arcsec$): $\sim 99.96\%$
($\sim3.5\sigma$)

 \item outer sample ($100\arcsec<r<360\arcsec$): $\sim 85.5\%$
($\sim1.5\sigma$)

\end{enumerate}

These results confirm the existence of a significant dip in the
radial distribution worthy of further discussion.

To see this effect in a different way, we have computed the doubly
normalized ratios for the BSS and the RGB stars, following the
definitions made by F93:

\begin{displaymath}
R_{\rm BSS} = {{(N_{\rm BSS}/N_{\rm BSS}^{\rm tot})} \over 
{(L^{sample}/L_{tot}^{sample})}} 
\end{displaymath}

and

\begin{displaymath}
R_{\rm RGB} = {{(N_{\rm RGB}/N_{\rm RGB}^{\rm tot})} \over 
{(L^{\rm sample}/L_{\rm tot}^{\rm sample})}} 
\end{displaymath}

\noindent
respectively.

The numbers of BSS and RGB in each annulus, the sampled luminosity,
and the resulting ratios with the associated errors are reported in
Table 5. The {\it relative frequency} of BSS so obtained is then
plotted as a function of the distance from the cluster center in
Figure 8 and compared with the corresponding one for the RGB
``reference'' stars. Note that an additional annulus with
$360\arcsec<r<600\arcsec$ has been added using the candidates by
Sandage (1953) to compute $R_{\rm BSS}$. This annulus gives at least an
indication of the BSS population in the outermost parts of the
cluster, even though we have no data to compute the corresponding
$R_{RGB}$ value.

As can be seen, the BSS specific frequency reaches its maximum at the 
center of the cluster, showing no evidence of a BSS
depletion in the core of M3, contrary to the claim of BHS.
We stress that this conclusion has been obtained under 
the most conservative assumption that only the 32 \br\ BSS 
detected on the basis of the UV CMD (in the innermost bin) are real. 

\begin{table*}[tb]
\begin{center}
\label{t:bssvsrgb}
\caption[]{ Numbers and relative frequencies for BSS and RGB stars}
\arraycolsep=\tabcolsep
$\begin{array}{lcccccccc}
\hline\hline
\\
{\rm Annulus} & N_{\rm BSS} & N_{\rm RGB} & L^{\rm sampled}/10^4L_\odot 
& L^{\rm sampled}/L_{tot}^{\rm sampled} & R_{\rm BSS} & \epsilon_{\rm BSS}
& R_{\rm RGB} & \epsilon_{\rm RGB}\\
\\
\hline
\\
\\
0\arcsec-20\arcsec    & 32 & 193 & 4.0 & 0.10 & 2.76 & 0.27 & 1.20 & 0.12 \\
20\arcsec-50\arcsec   & 36 & 391 & 8.9 & 0.23 & 1.40 & 0.26 & 1.09 & 0.08 \\
50\arcsec-100\arcsec  & 22 & 376 & 9.4 & 0.24 & 0.81 & 0.31 & 1.00 & 0.08 \\
100\arcsec-150\arcsec &  3 & 220 & 5.7 & 0.14 & 0.18 & 0.67 & 0.96 & 0.09 \\
150\arcsec-210\arcsec &  3 & 162 & 4.6 & 0.12 & 0.23 & 0.67 & 0.88 & 0.09 \\
210\arcsec-290\arcsec & 10 & 128 & 4.2 & 0.11 & 0.82 & 0.41 & 0.76 & 0.09 \\
290\arcsec-360\arcsec &  8 & 111 & 2.6 & 0.07 & 1.06 & 0.45 & 1.06 & 0.13 \\
\\
\hline
\end{array}$
\end{center}
\end{table*}

\subsection{Dynamical simulations: comparison with the models}

The peculiar radial distribution of BSS in M3 has been the object of a 
detailed study by Sigurdsson \etal\ (1994).
They presented the results of a simulation of the dynamical evolution of 
a BSS population in a cluster with structural parameters similar to M3.
Sigurdsson \etal\ assumed that stellar collisions during
interaction in the core of the cluster between (primordial) binary 
and single stars are the dominant mechanism for
the M3 BSS formation. The normalized radial distribution for a sample 
of $\sim300$ BSS obtained from this simulation is overplotted to the
observed radial distribution in Figure 9.
As noted by Sigurdsson \etal\, the overall morphology of the 
simulated distribution is qualitatively identical to the observed one,
being able to reproduce:

\begin{enumerate}
\item the zone of avoidance at $\sim 5 r_c$

\item the rising BSS density for $r>8r_c$.
\end{enumerate}

This situation arises as follows: the outer BSS
have been formed in the core and then ejected into the outer regions
by the recoil from the interactions. Those binaries which get 
kicked out to $r < 7 r_c$ rapidly drift back to the center of the
cluster due to mass segregation, leading to a concentration of BSS near
the center and a paucity of BSS in the outer parts of this region.
More energetic kicks will take the BSS to larger distances; these stars 
require much more time to drift back toward the core and may account 
for the overabundance of BSS at large distance.

It is quite interesting to note that the simulated BSS density in 
the innermost radial bin is a sensitive function of the ratio of the BSS
lifetime to the halfmass relaxation time which, as noted by 
Sigurdsson \etal\ 1994, are highly uncertain.
In particular, they set this parameter to a very low value since BHS
observations at that time indicated  a low BSS density in the core.
The new \hst\ observations (GYBS, BPQ, and this paper) show a clear 
overdensity of BSS in the core. 

The observed bimodal distribution can provide interesting constraints.
Sigurdsson \etal\ (1994) suggested that it could arise 
either because the dynamical friction time-scale is short compared to BSS
lifetime or because the BSS do not receive significant kicks on formation.

\section{BSS Luminosity function}

Several authors (Bailyn 1992,1995, Auri\`ere \etal\ 1990, Fusi Pecci \etal\
1992,1993, Sarajedini 1993) have suggested that BSS with different
origin could have different photometric and spectroscopic
characteristics.  A potentially fruitful way to check these
possibilities is to study in detail the BSS LFs and, in particular,
its radial behavior in a cluster like M3 where a bimodal radial
distribution has been found.

In particular, we can carry out a {\it direct} test of the validity of the 
scenario presented by Bailyn and Pinsonneault (1995--BP95). They
computed evolutionary tracks of BSS generated by mergers of 
primordial binaries and stellar collisions, and presented theoretical LFs
for both scenarios, suggesting that BSS made from collisions should be 
systematically brighter than those made from mergers of primordial binaries.

Even in our first updated BSS catalog (Fusi Pecci \etal\ 1993), we
found quite convincing evidence that the BSS LFs for clusters with
$\log \rho_o<3$ and $\log \rho_o>3$, respectively, seem to be
different (at more than $3\sigma$). This earlier result supported the
hypothesis that the BSS formation mechanism varies with varying overall
cluster structural properties. It is thus quite easy to compare the
available BSS data with the LFs predicted by BP95.

Since the theoretical LFs computed by BP95 are given in bolometric luminosity,
a correction must be first applied to the data before comparing them with their
models. Assuming for the absolute magnitude of the TO level $M_V=4.0$ and
a differential Bolometric Correction equal to 0.1 mag between the BSS and 
the TO stars, following BP95, we computed for each star the quantity:
\begin{displaymath}
 Log(L/L_{\rm TO}) = 0.4(3.9-M_V)     
\end{displaymath}

\noindent In Figure 10 we compare the BSS LFs for high density clusters
({\it panel a)} and low density clusters ({\it panel b)}), respectively,
with the corresponding theoretical LFs.
The BSS LF for low-density clusters appears to be consistent
with that predicted from the merging of primordial binaries, while the BSS
LF for high-density clusters fits better to the claim that they were
mostly formed by collisions.  Though not conclusive---both the available
data and the models need to be improved---this comparison adds some
observational support to the claim that BSS formation mechanisms are
affected by environmental conditions. 

We can now test further this idea and the model proposed by BP95
by using the available BSS catalog in M3 to compare the BSS LFs obtained 
from different radial areas. In fact, as pointed out first in our
previous paper (F93), the bimodal BSS radial distribution in M3 
naturally leads one to imagine that the BSS located in the inner regions
of the cluster could have a different origin with respect to 
those populating the outer zones. 

Following this suggestion and using the BSS sample we published in
F93, BP95 found convincing evidence that the BSS in the outer regions
of M3 were probably formed from mergers of primordial binaries. They
were however unable to reach firm conclusions concerning the innermost
regions ($r<20\arcsec$), where the sample was very small and covered a
restricted luminosity range (in order to avoid the contaminations from
supra-SGB stars).  The \hst\ data give us the opportunity to further
investigate this aspect.

Therefore, we have obtained the LFs for the BSS (a) located within the
region $0\arcsec<r<100\arcsec$ from the cluster center (see Figure 8)
and with $B<18.6$, and (b) located within the annulus with
$210\arcsec<r<360\arcsec$.  Following BP95, we have adopted:
\begin{displaymath}
 Log(L/L_{TO}) = 0.4(19.1-V)     
\end{displaymath}

\noindent
and computed the LFs over luminosity bins of 0.1 ($=0.25$ mag).

In Figure 11 the observed BSS LFs in the two regions are compared to
the theoretical ones obtained from the BP95 models. The observed LF in
the inner region is clearly compatible with that predicted within the
collisional framework, while that obtained in the outer annulus is
consistent with the prediction of primordial binaries merging.

As noted above, for the comparisons of the BSS LFs of different
clusters, these indications are still too crude to allow us to draw
any firm conclusion based on a complete statistical analysis. However,
in our view this evidence confirms the possibility that two different
populations of BSS can actually be present in M3.

\subsection{Special warning}

Concerning the possible interpretation of the global BSS
properties (i.e. radial distribution and LF), we have to note that the 
comparison with the (still uncertain) theoretical models would
yield a possible contradicting scenario. In fact:

\begin{itemize}
 
\item (a) the radial distribution (see section 7.3) fits properly to 
the Sigurdsson \etal\ (1994) model, which suggests that all BSS are formed
via collisions in the central area and only a fraction 
of them are kicked out at large distance;

\item (b) the LFs seem to be in agreement with the BP95 models
which would lead to interpret the whole BSS population as formed by
two main mechanisms: collisions (in the inner regions) and merges 
of primordial binaries (in the outer regions).

\end{itemize}

We cannot solve the {\it dilemma} on the basis of the available data. 
Spectroscopic information could perhaps add useful hints on the issue.
In particular, data on the detailed chemical abundances (especially,
helium) and on rotational velocities could set important constraints
on modeling the BSS formation and evolution.

\section{Evolved BSS on the HB?}

Renzini and Fusi Pecci (1988) first suggested that some stars located at the 
red extreme of the HB could be {\it evolved} BSS, experiencing the core 
helium burning phase. Then, assuming a binary origin for the BSS (which, 
in turn, leads them to have high masses) and using the models computed by 
Seidl \etal\ (1987), Fusi Pecci \etal\ (1992) identified 47 red HB objects 
as possible post-BSS candidates in a sample of 10 GGCs.
In particular, Renzini and Fusi Pecci (1988) identified 7 stars in M3
slightly brighter and redder than {\it normal} HB stars which could
represent the evolved-BSS population.

To pursue this idea in our subsequent studies of M3, we (PH94, CCD97) 
defined a photometric box suitable to
identify these post-BSS candidates, labeled as ER (Extreme Red)-HB
stars. 
Now, based on the whole PHOTO+CCD+HST sample which 
we have secured, we estimate the total ER-HB population to be 19 stars.
For clarity, we have plotted a zoomed CMD of the HB in Figure 12, and the box
delimiting the ER stars in HB has been shown. 

Having defined a set of ``plausible'' post-BSS candidates, one can then 
check whether these stars have retained memory of the unusual radial 
distribution of their progenitors.

Unfortunately, the ER sample is {\it intrinsically} quite poor, so we
have defined only 4 radial bins instead of 7, and have computed the
double-normalized frequency (R$_{ER}$) as defined for the BSS:

\begin{displaymath}
 R_{\rm ER}= {{(N_{\rm ERanul}/N_{\rm ERtot})} \over {(L_{\rm
anul}/L_{\rm tot})}}  
\end{displaymath}

\noindent
Where $N_{\rm ERanul}, L_{\rm anul},~N_{\rm ERtot},~{\rm and}~L_{\rm
tot}$ are the numbers of detected ER-HB stars and the sampled
luminosity in each annulus, and the total sample respectively.

In Figure 13 {\it panel a}, the relative frequency of ER-HB stars in
M3 is plotted as a function of the radial distance from the cluster
center, while {\it panel b} shows the radial behavior of the BSS.  As
can be seen the overall trend is qualitatively the same, and the
region with $100\arcsec<r<200\arcsec$ characterized by a depletion of
BSS shows a clear deficit of ER-HB stars too.  The statistical
significance of this depletion is admittedly poor since only 3 stars have
been detected compared to the 10 expected in the region of the dip.
The observed number is only a
$\sim 2\sigma$ fluctuation from the expected value.  Clearly it is too
speculative to conclude that we have shown the existence of a
true connection between the ER-HB stars and the BSS.  On the other
hand, it may be useful for future analyses to keep track of such a
clear qualitative agreement between the two radial distributions.

If one assumes that the connection between the ER-HB stars and
the BSS is real, one can relate the
population ratios and the lifetimes of these evolutionary stages.
The ratio between the number of BSS and ER-HB stars 
using our sample in M3 ($r<360\arcsec$) is 
\begin{displaymath}
 {{N_{\rm BSS}}\over{N_{\rm ER-HB}}}={{122}\over{19}}=6.4  
\end{displaymath}

\noindent
which is in good agreement with a mean value of 6.6 over a sample 
of 10 GGCs found by Fusi Pecci \etal\ (1992). 

For comparison, Greggio \& Renzini (1990) give a relation for the expected 
number of ER-HB as a function of the sampled luminosity on the basis of 
the so-called fuel consumption theorem (Renzini and  Buzzoni 1986)
yielding:
\begin{displaymath}
 N_{\rm ER-HB}= 6 \times 10^{-13} L_T t_j    
\end{displaymath}

Assuming $t_{\rm HB}\sim 10^8$yr from Seidl \etal\ (1987) and
considering that with the complete PHOTO+CCD+HST survey we have
sampled $\sim4 \times 10^5\, L_{\odot}$ (i.e. $\sim 77 \%$ of the whole
cluster luminosity), we have that the predicted number of ER-HB stars
would be $N_{\rm ER-HB} \sim 24$, 
which is in excellent agreement with our observed sample (19 stars).
However, first, one has to note that the above relation for the lifetime
has actually been calibrated using the population of BSS and ER-HB 
in the outer regions of M3 (Buonanno \etal\ 1986), 
so this result cannot really be considered to be an independent 
confirmation of the connection between the two stellar populations. 
Second, the fuel consumption theorem may well not apply to BSS
straighforwardly, since they probably start out with considerable fuel 
already burned in their cores.

Despite the care we have taken, the connection of ER-HB stars with
the post-BSS is ``suggestive'' rather than proven. 
We think it is  worthwhile to  pursue the detailed search and study of 
the post-BSS since  they {\it must} be detectable {\it somewhere} in the CMD
of any GGC containing BSS (Renzini and Fusi Pecci 1988, Fusi Pecci
\etal\ 1992). The HB is still the most promising
location where they could be identified as (if truly more massive)
they could somehow differ from ``normal'' cluster stars.  On the other
hand, if correct, our previous estimate indicates that to detect 1
post-BSS during the core helium burning phase (ER-HB), one must sample
$\sim 2 \times 10^4\, L_{\odot}$. This implies that in order 
to obtain a statistically significant sample of ER-HB stars one 
should sample large fractions ($>50 \%$) of the brightest GGCs.

Given the difficulty of observing samples adequately large to connect 
the ER-HB and post-BSS
via population studies, it might be appropriate to consider alternate
techniques. In either the merged binary or collision mechanisms one
might expect a substantial amount of CNO processed material to end up
in the surface layers of the stars. This material could show the same
kind of abundance anomalies found by Kraft \etal\ (1996 and references
therein) in
highly mixed red giants, i.e., extreme oxygen depletion coupled with
sodium and aluminum enhancements. A differential study comparing 
ER-HB/post-BSS candidates to normal RHB stars could be quite revealing. 

\section{Variable BSS in M3}
Mateo \etal\ (1996 and references therein) have shown the existence of 
variable BSS and classified them as SX Phoenicis, eclipsing binaries or 
contact binaries W UMa. 

Our observations do not have a time coverage suitable to look for
variability. It is thus difficult to select BSS variable candidates
only on the basis of our photometry.
Since the U exposures cover $\sim$25 minutes in time duration,
possible BSS candidates can be perhaps found by looking for particularly large
values of $\sigma_{U}$, the rms frame-to-frame scatter.

The mean $\sigma_{U}$ in the range of magnitude 17.5$<$U$<$18.5
for the RGB stars and for BSS is $\sigma_{U}^{RGB}$=0.02.
It is quite interesting to note that BSS/5866 shows a $\sigma_{U} \sim
0.16$ which is $\sim 8\sigma$ greater than the average value; this BSS
has not been catalogued as possible variable by GYBS, who, on the other hand,
found only one variable BSS in their sample: BSS/576 (in their catalog).
For this star they published also a
possible light curve (see figure 16 in GYBS), however 
in our photometry this star --BSS/1498-- has $\sigma_{U}$=0.033.
It is thus hard to say any additional comment on variability.

\section{Three very bright BSS in M3}

Figure 14 shows the CMD in the ultraviolet plane with the total sample
of BSS identified using different symbols ({\it big dots} $=$ bright
BSS, and {\it asterisks} $=$ faint BSS).  Overplotted on the data are
two isochrones (2 and 14\,Gyr) computed by Dorman (1995, unpublished).
These respectively have a TO-mass of $\sim 0.8$ and $\sim 1.5\,M_{\odot}$.
Most of the BSS candidates are consistent with a mass $<2 M_{\rm
TO}^{\rm M3}$, and are thus clearly compatible with the hypothesis that
they are the result of advanced stages of evolution of binary systems.

However, it is also quite interesting to note that there are three objects 
which are $\sim0.5\,$mag brighter than the predicted TO of the 1.5\msun\
track. None of these (namely No. 24768, 50012 and 6674) were present
in the selection by BHS and GYBS. Two (No. 24768 and 6674) are 
are actually brighter (0.3 mag) than the BSS box defined by BHS, and
are located slightly fainter than the blue region of the HB.
No. 6674 is located within the field covered by GYBS, but it was
not considered to be a BSS candidate.
The third object (No. 50012) is an HB star in BHS and F93 while it is outside 
the field covered by GYBS. Since this object is positioned in our frames
very close to the WF2 edge, we cannot exclude the existence of a sizable
error in our photometry, and will thus exclude it from the following analysis.

The positions of these stars in the UV-CMD suggest that they could be
bright BSS rather than ``peculiarly faint'' HB stars, since the HB is
quite well defined and these stars are located $\sim0.5\,$mag below 
the lower
envelope of the HB (the plausible Zero Age HB).  Since they do not
show any variability, they cannot be identified as candidate RRLyrae
variables observed out-of-phase in the two colours. If they are
``true'' BSS, these objects could have originated from multiple systems
(3 or more objects), and would demonstrate that such complex
interactions can take place in the core of M3.

The possible formation of multiple systems has been proposed by 
Leonard (1993) from the interaction between two binary systems. Their
existence could, for instance, explain the BSS F81 in M67, which shows 
a mass greater than twice the TO mass (Leonard and Linnel 1992).
The predicted percentage of the ``stable'' multiple systems from collision 
of binary systems is $\sim2$--4\% (Leonard 1996), fully compatible 
with our observed percentage ($3/171\sim2$\%).

Alternatively, the very high luminosity reached by the brightest BSS
could be ascribed to the effects of helium mixing in the envelope
during the collision. In particular, Sandquist \etal\ (1997) describe
the evolution of massive objects obtained from hydrodynamical
simulations of direct collisions of single and binary stars. They
suggested that the luminosities of the brightest BSS (much brighter
than expected for a $2\,M_{\rm TO}$ star) can still be explained by
the merger of two $1\,M_{\rm TO}$ stars. The resulting object has
overall helium enrichment because the collision mixes the helium
produced during the prior evolution.

\section{Conclusions}

Using new \hst\ observations and a thorough  revision of all 
previous surveys of
blue straggler stars in M3, we carried out 
a complete re-analysis of BSS properties and, in particular,
of their radial distribution. The results are:

\begin{itemize}

\item  From the careful revision of all the 263 BSS candidates
proposed at least by one of the various surveys carried out so far, we
have adopted an updated catalog of the BSS candidates in M3 which includes 
122 {\it bright} (with $B<18.6$, $V<18.3$, $m_{255}<19.0$) and 141
{\it faint} objects. The {\it bright global} sample within the radial
region at $0\arcsec<r<360\arcsec$ includes 114 objects.  This
sample, hopefully complete over the whole region, has been 
used for the subsequent analysis. All the candidate objects
are listed along with comments in Table 4. The {\it faint} sample is
surely incomplete and it may be strongly biased by the
selection criteria which are hard to define at the separation 
between the BSS region and the MS stars.

\item  The BSS radial distribution is clearly peaked at the cluster center.
BSS candidates in the inner regions (see section 7.1) are more 
concentrated than ``reference'' normal SGB and RGB stars.
In particular, star counts yield a rather high specific frequency 
for the BSS in the cluster core, contrary to previous claims
which were affected by incomplete samples.

\item  The global radial distribution of the BSS in M3 is bimodal,
with a clear-cut dip in the region at $100\arcsec<r<200\arcsec$ and 
a rising 
trend in the outer region (out to $r\sim360\arcsec$), as first noted 
by F93. This evidence adds support to the idea that
different mechanisms of BSS formation and survival could be at work
also within the same cluster, or that special segregation effects
take place during the dynamical cluster evolution. 
There is an interesting  {\it qualitative} agreement
between the observed radial distribution of the {\it bright} BSS and
the predictions of the dynamical models computed by Sigurdsson
\etal\ (1994) which take into account both star collisions and
merging of primordial binaries.

\item  The total Luminosity Function is consistent with that obtained
for all the known BSS candidates (about 700 in $\sim 35$ GGCs, see Fusi Pecci 
\etal\ 1992, Sarajedini 1993, Bailyn 1995). Due to quoted selection biases,
the faint end of the LF is still highly uncertain. In addition, as
already found in other clusters (see references above), there are a
few BSS candidates which seem too bright to possibly have originated
from  binary systems.

\item  The LFs of the inner ($r<100\arcsec$) and outer ($r>210\arcsec$) regions
are different. The brightest BSS are slightly more frequent in the central 
zones. In particular, comparing with the available models (BP95),
one finds that the BSS LF in the outer regions $(r>210\arcsec)$ is consistent
with the scenario based on merging of primordial binaries as basic
formation mechanism. In the central region, the BSS LF is 
compatible with  the collisional origin, though the significance
of the difference is small.

\end{itemize}

From the above considerations, the BSS content in M3 can be 
schematically described as follows:

\begin{enumerate}

\item A large fraction of the BSS populate the very central 
cluster regions. The LF suggests that most 
of them were probably generated by stellar 
collisions (possibly) involving primordial binaries, which are
actually disrupted in the core (BP95).

\item The dip detected in the radial distribution is probably due
to mass segregation. In particular, according to Sigurdsson \etal\
(1994), primordial binaries lying less than $ 7 r_c$ from the cluster
center are attracted toward the center and then disrupted, and BSS
produced by collisions in the core and kicked out at $r<7\arcsec$ 
drift back to the core in a short time-scale.

\item  On the other hand, most of the BSS in the outer regions 
are originated 
from the merging of primordial binaries, though a few of them
could also be originated by collisions in the core and pushed out
at large distances because of very energetic dynamical kicks
(Sigurdsson \etal\ 1994).

\end{enumerate}

A deeper discussion of the BSS properties in this and in other GGCs
will be surely possible as soon as new data will become available 
from the \hst\ frames currently taken or reduced for
the central regions of many Galactic globular clusters.

\begin{acknowledgements}
We gratefully thank 
the referee, Charles Bailyn, for helpful comments and suggestions to
improve the presentation, and
G. Iannicola and I. Ferraro whose expertise on
ROMAFOT was invaluable for the completion of this work.
This research was supported by the {\it Ministero 
dell' Universit\`a e della Ricerca Scientifica e Tecnologica }
(MURST), the {\it Agenzia Spaziale Italiana} (ASI), and the
{\it Gruppo Nazionale di Astronomia del Consiglio Nazionale delle
Ricerche} (GNA-CNR). R.T.Rood \& B.Dorman are 
supported in part by NASA Long Term
Space Astrophysics Grant NAGW-2596.
\end{acknowledgements}

\begin{figure}[htbp]
\psfig{figure=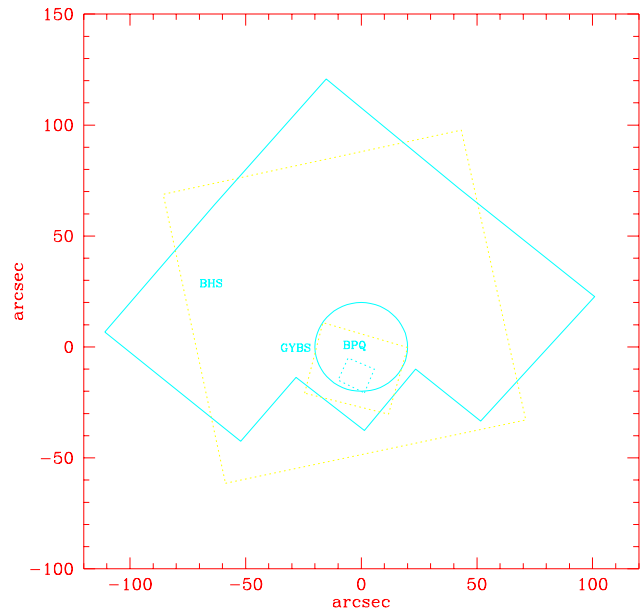,width=8.8cm,clip=}
\caption{ The areas surveyed in various BSS studies of M3. The global
(PC+WF2+WF3+WF4) field of view of the \hst\ field is shown by the solid 
line.  The dotted squares represent the boundaries of previous surveys in the
center of M3: Bolte \etal\ (1993, labeled BHS), Guathakurta \etal\ (1995,
label GYBS), Burgarella \etal\ (1995, labeled BPQ), respectively.  The 
circle at $r=20\arcsec$ is the inner limit of our previous
ground-based surveys (F93, CCD97).}
\label{f:f1}
\end{figure}

\begin{figure}[htbp]
Fig. 2 Converted to .jpg file
\caption{ UV-CMD in the plane ($m_{255},~m_{255}-U$) for more than 
18,000 stars
identified in the \hst\ field. The variable stars are not shown. 
}
\label{f:f2}
\end{figure}

\begin{figure}[htbp]
Fig.3 converted to .jpg file
\caption{ 
Blue stragglers in M3. The 
solid line at $m_{255}=19.4$ is the assumed limiting magnitude
(which is $\sim 5\sigma$ above the TO level).
The dashed line at $m_{255}=19.0$ 
(which corresponds to $B\sim18.6$  in PH94)
divides the total sample in two sub-samples:
the Bright sample (Bright-BSS){\it solid circles}
and the the Faint sample (Faint-BSS){\it asterisks}
}
\label{f:f3}
\end{figure}

\begin{figure}[htbp]
Fig. 4 converted to .jpg file
\caption{
The BSS candidates (filled triangles) in different parts of the cluster, 
and the RGB (dots) population for comparison.
{\it Panel (a):} the $(V,~m_{255}-V)$ CMD for the region covered by \hst -- 
the two lines indicate the limiting magnitudes for
the RGB stars ($B<18.6$) and the BSS ($m_{255}<19.0$); 
{\it panel (b):} the $(V,~V-I)$ CMD for the region with $r<210\arcsec$ outside
the \hst\ field; 
and {\it panel (c):} the $(V,~B-V)$ CMD for the outer
regions (to r$<360\arcsec$). In panel (c) the empty squares are 7 bright BSS
from Sandage (1953) with $r>360\arcsec$:  their original colours have been
shifted ($\Delta V=0.077$ and $\Delta (B-V)=0.15$) in order to fit to the BSS
sequence. }
\label{f:f4}
\end{figure}

\begin{figure}[htbp]
\psfig{figure=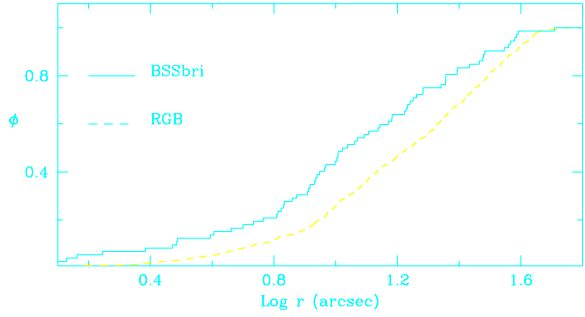,width=8.8cm,clip=}
\caption{
Cumulative distribution of the Bright-BSS (full line) and 
the RGB stellar population
(dashed line) in the \hst\ field. 
}
\label{f:f5}
\end{figure}

\begin{figure}[htbp]
\psfig{figure=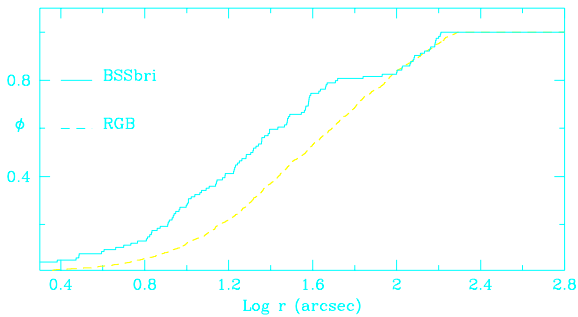,width=8.8cm,clip=}
\caption{
Same as Figure 5 over the distance range $0-6'$. 
}
\label{f:f6}
\end{figure}

\begin{figure}[htbp]
\psfig{figure=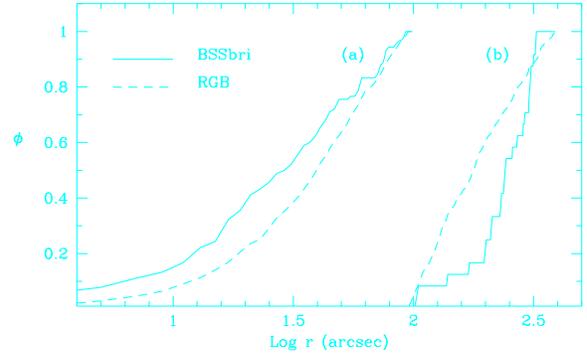,width=8.8cm,clip=}
\caption{
Same as Figure 5 
for the sub-sample of stars at (a) $r<2.5'$ and (b) $2.5'<r<6'$.
}
\label{f:f7}
\end{figure}

\begin{figure}[htbp]
\psfig{figure=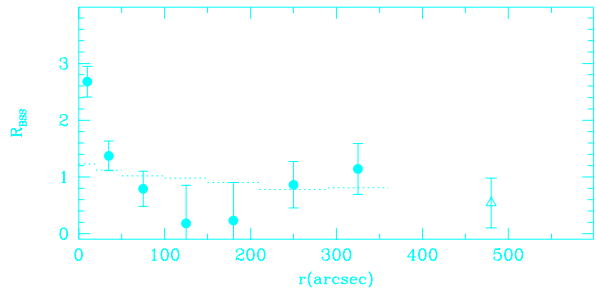,width=8.8cm,clip=}
\caption{
The relative frequency of BSS in M3 is plotted as a function of the radial 
distance from the cluster center. The horizontal lines show the relative 
frequency of the RGB stars used as a comparison population.
Note that for $r>6'$ only the relative frequency of BSS has been 
computed using the  Sandage (1953) candidates.
}
\label{f:f8}
\end{figure}

\begin{figure}[htbp]
\psfig{figure=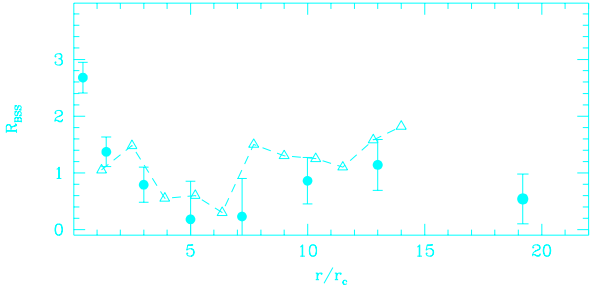,width=8.8cm,clip=}
\caption{
The relative frequency of BSS in M3 is plotted as function of the radial 
distance from the cluster center in units of core radii.
The frequency expected from the simulation by Sigurdsson \etal\ (1994) has been
overplotted as empty triangles.
}
\label{f:f9}
\end{figure}

\begin{figure}[htbp]
\psfig{figure=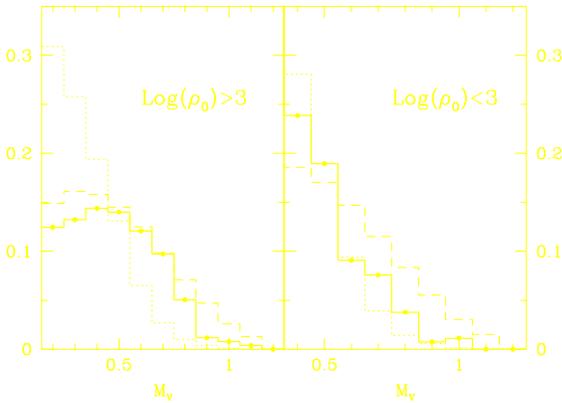,width=8.8cm,clip=}
\caption{The BSS luminosity functions (solid lines)
for clusters having $\log(\rho_o)$  greater and smaller than 3 
(panels (a) and (b), respectively) 
are compared to the theoretical LFs  for collisional BSS 
(dashed line) and primordial binary merger BSS (dotted line).
The data are from Fusi Pecci \etal (1993), the theoretical LFs are from 
BP95. }
\label{f:f10}
\end{figure}

\begin{figure}[htbp]
\psfig{figure=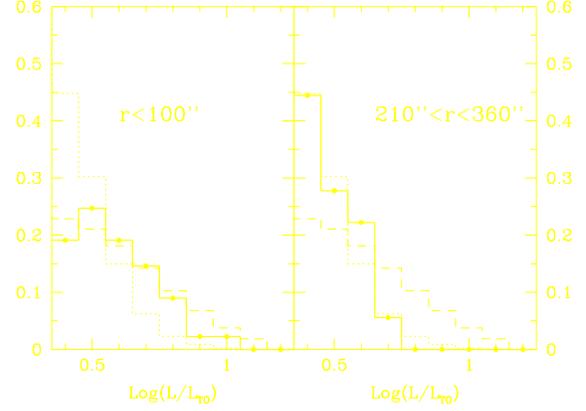,width=8.8cm,clip=}
\caption{The BSS luminosity functions for the inner regions ($r<100\arcsec$ 
-- panel (a)) and outer regions ($r>210\arcsec$ -- panel (b))
are compared to the theoretical predictions: line symbols have
the same meaning as in Figure 9.
}
\label{f:f11}
\end{figure}

\begin{figure}[htbp]
\psfig{figure=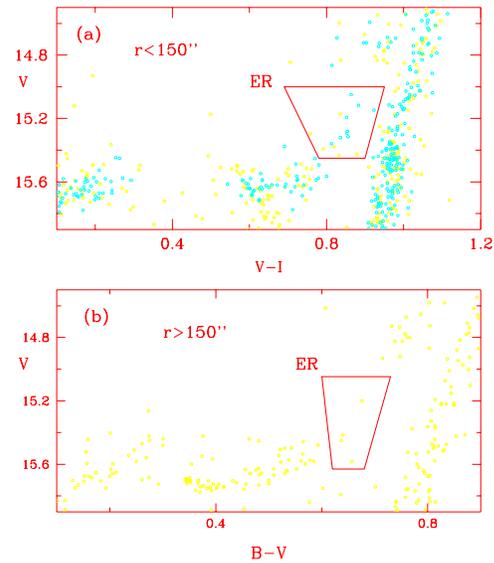,width=8.8cm,clip=}
\caption{ Zoomed CMD of the Horizontal branch region. 
The box defines the area in
which the BSS progeny are expected (see PH94, CCD97). 
{\it Panel (a)}: ($V,~V-I$) CMD for stars at $r<150\arcsec$; 
{\it panel (b)}: ($V,~B-V$) CMD for stars at $150\arcsec<r<360\arcsec$ and 
outside the \hst\ field.  }
\label{f:f12}
\end{figure}

\begin{figure}[htbp]
\psfig{figure=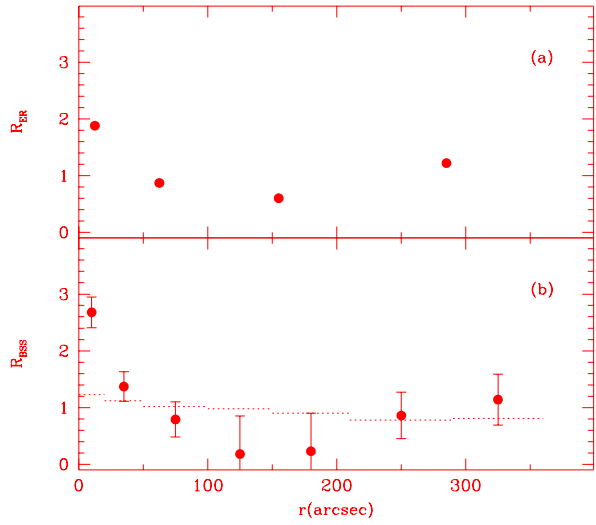,width=8.8cm,clip=}
\caption{
{\it Panel (a)}
The relative frequency of ER-HB stars in M3 is plotted as 
function of the radial 
distance from the cluster center.
{\it Panel (b)} The relative frequency of the BSS
is plotted for comparison.
As can be seen the overall trend for the two types of star 
is qualitatively the same.
}
\label{f:f13}
\end{figure}

\begin{figure}[htbp]
\psfig{figure=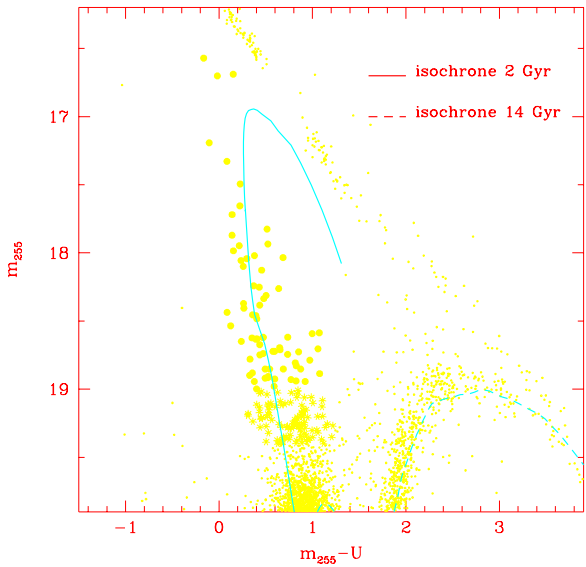,width=8.8cm,clip=}
\caption{
Zoomed CMD of the BSS region: the  symbols have
the same meaning as in Figure 3.
Two isochrones have been plotted: 
{\it (i)}  14 Gyr (dashed line)  with a TO mass of 
$\sim 0.8\msun$ which nicely fits to the TO region and the RGB; and 
{\it (ii)} 2 Gyr (solid line) which corresponds to a TO mass
of $\sim 1.5\msun \sim 2 \times M_{TO}$ (Dorman 1995, unpublished). 
}
\label{f:f14}
\end{figure}

\clearpage

{
\hsize=15cm \vsize=20cm \hoffset -1cm

\begin{center}
\pagestyle{empty}

\small
{\bf Tab.2 HST BSS : bright sample} 
\vskip0.6truecm
\begin{tabular}{rrrrrrrr}
\hline
 \multicolumn{1}{c}{N} &\multicolumn{1}{c}{V} &\multicolumn{1}{c}{I}
&\multicolumn{1}{c}{U} &\multicolumn{1}{c}{$m_{255}$} 
&\multicolumn{1}{c}{X} &\multicolumn{1}{c}{Y} &\multicolumn{1}{c}{r}\\ 
\hline
7821 & 18.176 & 17.748 & 18.389 & 18.979&  -1.040&   0.610&   1.206\\
5866 & 17.863 & 17.441 & 18.151 & 18.622&  -1.750&   1.230&   2.139\\
5906 & 17.510 & 17.303 & 17.841 & 18.100&  -2.680&  -0.190&   2.687\\
7785 & 17.963 & 17.820 & 18.271 & 18.624&  -1.110&  -2.690&   2.910\\
6674 & 16.593 & 16.507 & 16.735 & 16.571&   2.210&   2.740&   3.520\\
6430 & 17.310 & 16.677 & 17.809 & 18.886&  -1.930&  -4.430&   4.832\\
5981 & 18.016 & 17.435 & 18.311 & 18.748&  -2.750&  -5.260&   5.935\\
3736 & 17.801 & 17.650 & 18.096 & 18.455&   6.050&   1.060&   6.142\\
7678 & 17.028 & 16.941 & 17.243 & 17.328&  -6.060&  -0.870&   6.122\\
6318 & 18.166 & 17.984 & 18.412 & 18.650&  -1.040&   7.790&   7.859\\
7850 & 18.308 & 18.006 & 18.411 & 18.535&  -3.810&  -7.120&   8.075\\
6979 & 17.557 & 17.436 & 17.831 & 17.985&   9.070&  -1.470&   9.188\\
5313 & 17.637 & 17.005 & 18.022 & 18.944&   9.670&  -2.620&  10.019\\
6352 & 17.799 & 17.552 & 18.109 & 18.371&   9.720&  -4.800&  10.841\\
7069 & 17.284 & 16.953 & 17.641 & 18.020&   1.940& -11.490&  11.653\\
2926 & 17.082 & 16.970 & 17.297 & 17.192&  -8.150&  -9.960&  12.870\\
1769 & 16.989 & 16.448 & 17.350 & 18.035&  -0.940& -12.960&  12.994\\
47750 & 17.734 & 17.860 & 17.821 & 18.056&  -1.030&  13.280&  13.320\\
47934 & 18.114 &  ... & 18.311 & 18.852&  10.070&   9.090&  13.566\\
2444 & 17.655 & 17.035 & 17.991 & 18.807&  -9.140& -10.120&  13.636\\
1855 & 18.133 & 17.637 & 18.426 & 18.919&  -7.430& -12.430&  14.481\\
48425 & 18.294 & 17.995 & 18.393 & 18.915&   9.640&  11.420&  14.945\\
20075 & 17.516 & 17.174 & 17.810 & 18.313& -12.860&   9.880&  16.217\\
45876 & 17.691 &  ... & 17.948 & 18.384&  15.970&   3.520&  16.353\\
1244 & 18.451 & 18.202 & 18.566 & 18.944&  -0.120& -16.510&  16.510\\
3407 & 17.434 & 17.251 & 17.732 & 17.948& -11.740& -12.370&  17.054\\
48119 & 17.657 & 17.166 & 18.002 & 18.829&   8.490&  15.030&  17.262\\
2922 & 17.731 & 17.242 & 18.011 & 18.745& -17.310&  -1.870&  17.411\\
7698 & 17.426 & 16.924 & 17.671 & 18.127&  11.840& -13.220&  17.747\\
1441 & 17.918 & 17.305 & 18.266 & 18.739&  -6.750& -17.110&  18.393\\
519 & 17.908 & 17.575 & 18.240 & 18.633& -16.200&  -9.070&  18.566\\
1498 & 16.918 & 16.534 & 17.312 & 17.826&  -8.610& -17.960&  19.917\\
4032 & 17.857 & 17.299 & 18.128 & 18.723&  15.340& -13.080&  20.159\\
5214 & 17.867 & 17.272 & 18.165 & 18.853&  19.200&  -6.880&  20.395\\
47269 & 18.005 & 17.657 & 18.146 & 18.619&   3.760&  20.010&  20.360\\
46285 & 17.879 & 17.762 & 18.058 & 18.711&  18.550&   9.900&  21.026\\
21131 & 17.908 & 17.458 & 18.160 & 18.930& -17.040&  13.700&  21.864\\
35187 & 17.668 & 17.354 & 17.858 & 18.336&  -8.090&  21.600&  23.065\\
\end{tabular} 
\end{center}
\newpage
\small
\begin{center}
{\bf Tab.2 {\it (continued)}} 
\vskip0.6truecm
\begin{tabular}{rrrrrrrr}
\hline
 \multicolumn{1}{c}{N} &\multicolumn{1}{c}{V} &\multicolumn{1}{c}{I}
&\multicolumn{1}{c}{U} &\multicolumn{1}{c}{$m_{255}$} 
&\multicolumn{1}{c}{X} &\multicolumn{1}{c}{Y} &\multicolumn{1}{c}{r}\\ 
\hline
21018 & 18.014 & 17.439 & 17.432 & 17.655& -22.850&   5.490&  23.500\\
47081 & 17.998 & 17.820 & 18.224 & 18.631&   8.760&  23.160&  24.761\\
3189 & 17.887 & 17.536 & 18.146 & 18.723& -11.070& -23.040&  25.561\\
1534 & 17.398 & 17.100 & 17.745 & 18.042&   0.470& -27.420&  27.424\\
35335 & 17.864 & 17.597 & 18.080 & 18.484&  -1.940&  27.760&  27.828\\
48164 & 17.437 & 16.808 & 17.645 & 18.704&  28.250&   9.000&  29.649\\
30784 & 18.167 &  ... & 18.351 & 18.437&  -3.720&  30.190&  30.418\\
21270 & 17.516 & 17.583 & 17.732 & 17.871& -30.470&  -1.410&  30.503\\
47470 & 17.609 & 17.472 & 17.822 & 18.252&  32.350&  -8.320&  33.403\\
35310 & 18.283 & 17.954 & 18.524 & 18.881&   6.870&  33.190&  33.894\\
33784 & 17.555 & 17.408 & 17.870 & 18.241&  -5.540&  33.750&  34.202\\
50012 & ... & ... & 16.537 & 16.689 & -30.240 & -17.430 & 34.904\\
23174 & 17.934 &  ... & 18.238 & 18.673& -30.510&  18.460&  35.660\\
47022 & 17.045 & 17.109 & 17.267 & 17.494&  36.470&   2.750&  36.574\\
45819 & 17.305 & 16.628 & 17.595 & 18.592&  26.630&  27.660&  38.396\\
23518 & 18.195 & 17.645 & 18.359 & 18.853& -31.800&  21.410&  38.336\\
23211 & 17.078 & 16.819 & 17.412 & 17.935& -43.830&   2.730&  43.915\\
23745 & 17.838 & 17.187 & 18.116 & 18.940& -44.550&   8.730&  45.397\\
44630 & 18.461 & 16.755 & 18.571 & 18.899&  40.320& -21.150&  45.530\\
34704 & 17.675 & 17.034 & 17.874 & 18.726&  13.930&  43.360&  45.543\\
31956 & 17.443 & 16.752 & 17.816 & 18.787&   7.940&  48.910&  49.550\\
24768 & 16.628 & 16.553 & 16.720 & 16.702& -38.720&  30.970&  49.582\\
34147 & 18.107 &  ... & 18.330 & 18.926&   3.410&  54.180&  54.287\\
32140 & 17.340 & 16.859 & 17.624 & 18.262&   2.950&  58.400&  58.474\\
20674 & 17.843 &  ... & 18.046 & 18.695& -50.870& -31.580&  59.875\\
25464 & 18.326 & 17.969 & 18.596 & 18.998& -55.240&  24.130&  60.280\\
24688 & 18.202 &  ... & 18.448 & 18.780& -60.880&   2.860&  60.947\\
21863 & 18.075 &  ... & 18.382 & 18.908& -61.720& -33.870&  70.403\\
23457 & 17.870 & 17.718 & 18.140 & 18.406& -68.590& -24.110&  72.704\\
24338 & 17.667 & 17.139 & 17.884 & 18.619& -71.860& -16.300&  73.685\\
26508 & 17.824 & 17.413 & 18.066 & 18.460& -63.710&  41.630&  76.105\\
35060 & 16.856 & 16.132 & 17.513 & 18.586&  -5.500&  76.770&  76.967\\
26142 & 17.288 & 17.309 & 17.579 & 17.719& -76.680&  14.610&  78.059\\
26972 & 17.739 & 17.116 & 17.979 & 18.853&-101.480&  11.820& 102.166\\
\end{tabular} 
\end{center}
}

\clearpage

{
\hsize=15cm \vsize=20cm \hoffset -1cm

\begin{center}
\pagestyle{empty}

\small
{\bf Tab.3 HST BSS : faint sample} 
\vskip0.6truecm
\begin{tabular}{rrrrrrrr}
\hline
 \multicolumn{1}{c}{N} &\multicolumn{1}{c}{V} &\multicolumn{1}{c}{I}
&\multicolumn{1}{c}{U} &\multicolumn{1}{c}{$m_{255}$} 
&\multicolumn{1}{c}{X} &\multicolumn{1}{c}{Y} &\multicolumn{1}{c}{r}\\ 
\hline
   7570 &   18.685 &   18.146 &   18.882 &   19.192 &   -2.280 &   -1.060 &    2.514\\
   7262 &   17.931 &   17.366 &   18.233 &   19.120 &   -0.400 &   -4.370 &    4.388\\
   7613 &   17.581 &   16.769 &   18.026 &   19.128 &   -4.620 &   -0.680 &    4.670\\
   6832 &   18.509 &   18.104 &   18.749 &   19.372 &   -4.340 &   -4.520 &    6.266\\
   7339 &   18.072 &   17.794 &   18.418 &   19.286 &    6.820 &   -1.180 &    6.921\\
   6837 &   18.434 &   17.881 &   18.697 &   19.204 &   -5.230 &   -5.010 &    7.242\\
   6910 &   18.509 &   18.166 &   18.767 &   19.271 &    7.920 &   -0.390 &    7.930\\
   7879 &   18.318 &   17.934 &   18.547 &   19.103 &    4.330 &   -8.280 &    9.344\\
   7365 &   18.475 &   18.175 &   18.735 &   19.168 &   -2.190 &   -9.510 &    9.759\\
   6977 &   17.813 &   17.113 &   18.141 &   19.278 &   -7.900 &   -5.990 &    9.914\\
  47941 &   18.225 &    ... &   18.410 &   19.340 &    8.760 &    4.970 &   10.072\\
   8009 &   17.866 &   17.602 &   18.225 &   19.059 &    5.340 &   -8.590 &   10.115\\
   6902 &   18.027 &   17.477 &   18.458 &   19.248 &   -2.350 &    9.900 &   10.175\\
   5229 &   18.376 &   17.880 &   18.614 &   19.148 &   -2.910 &   11.000 &   11.378\\
   7068 &   18.432 &   18.148 &   18.663 &   19.353 &    2.000 &  -11.370 &   11.545\\
  46237 &   17.860 &   17.447 &   18.090 &   19.198 &    8.170 &    8.190 &   11.568\\
  47770 &   18.540 &   17.699 &   18.446 &   19.115 &    5.560 &   10.630 &   11.996\\
   2751 &   18.213 &   17.853 &   18.533 &   19.289 &   -2.770 &  -12.030 &   12.345\\
   5081 &   18.063 &   17.455 &   18.369 &   19.214 &    4.390 &  -12.040 &   12.815\\
  50082 &    ... &    ... &   18.253 &   19.299 &   -9.230 &    9.280 &   13.089\\
  47144 &   18.450 &   17.668 &   18.657 &   19.181 &   13.380 &    1.520 &   13.466\\
   5681 &   18.188 &   17.659 &   18.444 &   19.257 &    3.170 &  -13.540 &   13.906\\
  51933 &    ... &    ... &   18.347 &   19.254 &  -11.810 &    7.990 &   14.259\\
   1292 &   18.275 &   17.609 &   18.566 &   19.018 &  -11.680 &   -8.570 &   14.487\\
   7321 &   18.285 &   17.999 &   18.613 &   19.255 &   13.520 &   -6.350 &   14.937\\
   4925 &   18.522 &   18.057 &   18.767 &   19.216 &   11.040 &  -10.640 &   15.333\\
  51221 &    ... &    ... &   18.333 &   19.097 &  -14.700 &    4.550 &   15.388\\
  48269 &   17.614 &   16.900 &   17.974 &   19.081 &   16.530 &    0.380 &   16.534\\
  30161 &   18.600 &    ... &   18.743 &   19.393 &   -3.600 &   16.210 &   16.605\\
  20025 &    ... &    ... &   18.630 &   19.348 &   -8.020 &   14.750 &   16.789\\
   1398 &   18.017 &   17.349 &   18.360 &   19.283 &   -7.410 &  -15.280 &   16.982\\
   1284 &   18.162 &   17.680 &   18.400 &   19.107 &   -1.040 &  -17.420 &   17.451\\
  47970 &   18.284 &   17.545 &   18.288 &   19.185 &   17.530 &    0.070 &   17.530\\
  47973 &   18.222 &   17.703 &   18.326 &   19.240 &   17.580 &   -0.480 &   17.587\\
  48037 &   18.523 &    ... &   18.474 &   19.346 &   13.170 &   11.830 &   17.703\\
   3049 &   18.250 &   17.633 &   18.506 &   19.393 &  -17.520 &   -2.680 &   17.724\\
  43559 &   18.333 &    ... &   18.869 &   19.180 &    4.150 &   17.600 &   18.083\\
   2910 &   18.319 &   18.014 &   18.594 &   19.028 &    3.810 &  -17.880 &   18.281\\
\end{tabular} 
\end{center}

\newpage
\begin{center}
\pagestyle{empty}

\small
{\bf Tab.3 {\it (continued)}} 
\vskip0.6truecm
\begin{tabular}{rrrrrrrr}
\hline
 \multicolumn{1}{c}{N} &\multicolumn{1}{c}{V} &\multicolumn{1}{c}{I}
&\multicolumn{1}{c}{U} &\multicolumn{1}{c}{$m_{255}$} 
&\multicolumn{1}{c}{X} &\multicolumn{1}{c}{Y} &\multicolumn{1}{c}{r}\\ 
\hline
   2424 &   17.906 &   17.372 &   18.259 &   19.317 &    2.380 &  -18.420 &   18.573\\
  20538 &   18.200 &    ... &   18.286 &   19.170 &  -12.520 &   13.730 &   18.581\\
   4361 &   18.280 &   18.095 &   18.526 &   19.039 &   10.420 &  -15.900 &   19.010\\
  47581 &   18.273 &    ... &   18.441 &   19.381 &    5.420 &   18.310 &   19.095\\
  47541 &   17.913 &   17.374 &   18.073 &   19.059 &   19.350 &    3.100 &   19.597\\
  44938 &   18.368 &   17.963 &   18.575 &   19.336 &    7.690 &   18.610 &   20.136\\
  20753 &   18.141 &   17.665 &   18.310 &   19.201 &  -18.250 &    9.290 &   20.478\\
  48349 &   17.955 &    ... &   18.527 &   19.129 &   20.640 &   -0.900 &   20.660\\
   7547 &   18.191 &   17.891 &   18.397 &   19.023 &   16.140 &  -13.090 &   20.781\\
  35380 &   18.310 &    ... &   18.683 &   19.366 &  -10.200 &   18.400 &   21.038\\
  20133 &   18.235 &   17.769 &   18.420 &   19.239 &  -21.070 &   -0.170 &   21.071\\
  45855 &   18.204 &   17.762 &   18.351 &   19.352 &   19.420 &   10.660 &   22.153\\
   2979 &   18.270 &   17.640 &   18.593 &   19.120 &   -3.670 &  -22.590 &   22.886\\
  35354 &   18.514 &   17.777 &   18.660 &   19.224 &   -4.160 &   22.770 &   23.147\\
  20438 &   18.549 &    ... &   18.791 &   19.320 &  -24.150 &   -1.300 &   24.185\\
  48100 &   17.820 &   17.261 &   18.096 &   19.255 &    5.780 &   24.150 &   24.832\\
   3019 &   18.659 &   18.234 &   18.874 &   19.236 &  -19.130 &  -16.050 &   24.971\\
  48085 &   17.970 &   17.694 &   18.147 &   19.061 &   24.840 &    2.980 &   25.018\\
  48099 &   18.589 &   17.802 &   18.699 &   19.133 &    5.620 &   24.480 &   25.117\\
  47164 &   18.476 &    ... &   18.674 &   19.366 &   23.510 &   11.410 &   26.133\\
   2070 &   18.546 &   17.913 &   18.769 &   19.399 &   -4.520 &  -25.900 &   26.291\\
  45066 &   17.812 &    ... &   18.251 &   19.025 &   18.520 &   19.000 &   26.533\\
   1076 &   18.456 &   18.063 &   18.710 &   19.113 &  -22.970 &  -13.440 &   26.613\\
  47590 &   18.262 &   17.584 &   18.433 &   19.310 &   19.550 &   19.270 &   27.451\\
  34816 &   18.048 &   17.666 &   18.257 &   19.206 &   -8.470 &   26.700 &   28.011\\
  47529 &   18.214 &   17.723 &   18.433 &   19.088 &   23.260 &   16.730 &   28.652\\
  45165 &   17.853 &   17.242 &   18.145 &   19.200 &   25.820 &   12.810 &   28.823\\
  48012 &    ... &    ... &   18.554 &   19.278 &   11.080 &   27.160 &   29.333\\
  41178 &   17.949 &   17.606 &   18.163 &   19.008 &   27.300 &   10.890 &   29.392\\
  34446 &   18.776 &    ... &   18.882 &   19.385 &  -16.930 &   24.630 &   29.887\\
  34864 &   18.245 &   17.597 &   18.478 &   19.365 &   -0.330 &   30.150 &   30.152\\
  46418 &   17.956 &   18.311 &   18.144 &   19.350 &   30.290 &   -0.290 &   30.291\\
  33562 &   18.186 &   17.694 &   18.382 &   19.262 &  -16.480 &   26.930 &   31.572\\
  44185 &   18.208 &   17.690 &   18.408 &   19.370 &   32.150 &    5.060 &   32.546\\
  47485 &   18.380 &    ... &   18.363 &   19.203 &   16.460 &   29.680 &   33.939\\
  46932 &   18.051 &    ... &   18.218 &   19.149 &   11.710 &   31.870 &   33.953\\
\end{tabular} 
\end{center}

\newpage
\begin{center}
\pagestyle{empty}

\small
{\bf Tab.3 {\it (continued)}} 
\vskip0.6truecm
\begin{tabular}{rrrrrrrr}
\hline
 \multicolumn{1}{c}{N} &\multicolumn{1}{c}{V} &\multicolumn{1}{c}{I}
&\multicolumn{1}{c}{U} &\multicolumn{1}{c}{$m_{255}$} 
&\multicolumn{1}{c}{X} &\multicolumn{1}{c}{Y} &\multicolumn{1}{c}{r}\\ 
\hline
  33861 &   18.184 &    ... &   18.423 &   19.288 &   -2.260 &   36.600 &   36.670\\
  45828 &   18.311 &    ... &   18.543 &   19.376 &   28.260 &   24.840 &   37.625\\
  45007 &   18.447 &    ... &   18.652 &   19.187 &   23.660 &   29.860 &   38.097\\
  46434 &   18.695 &    ... &   18.841 &   19.279 &   39.330 &   -0.050 &   39.330\\
  21809 &   18.064 &   17.596 &   18.250 &   19.156 &  -40.500 &   -8.010 &   41.285\\
  23822 &   17.897 &   17.195 &   18.271 &   19.368 &  -28.660 &   29.830 &   41.367\\
  22805 &   18.356 &    ... &   18.436 &   19.260 &  -43.880 &   -2.030 &   43.927\\
  46857 &   17.838 &   17.295 &   18.095 &   19.100 &   42.060 &   13.900 &   44.297\\
  45472 &   17.711 &   17.046 &   18.051 &   19.268 &   44.490 &   -4.020 &   44.671\\
  23181 &   17.917 &   17.471 &   18.193 &   19.031 &  -45.260 &    0.750 &   45.266\\
  35241 &   18.212 &   17.743 &   18.411 &   19.222 &    9.500 &   45.370 &   46.354\\
  31833 &   18.447 &   18.098 &   18.616 &   19.366 &    5.110 &   48.290 &   48.560\\
  32122 &   18.426 &    ... &   18.606 &   19.275 &   17.710 &   45.610 &   48.928\\
  42948 &   18.494 &    ... &   18.570 &   19.394 &   45.670 &  -22.980 &   51.126\\
  43517 &   18.220 &   17.881 &   18.415 &   19.366 &   25.920 &   46.460 &   53.201\\
  34564 &   17.980 &   17.360 &   18.295 &   19.180 &  -25.850 &   47.500 &   54.078\\
  25134 &   18.219 &   17.576 &   18.418 &   19.261 &  -50.680 &   23.010 &   55.659\\
  44881 &   18.179 &   17.750 &   18.305 &   19.301 &   56.040 &  -28.440 &   62.844\\
  44584 &   18.257 &    ... &   18.499 &   19.284 &   65.090 &   13.220 &   66.419\\
  31898 &   18.183 &   17.650 &   18.409 &   19.325 &  -13.450 &   65.380 &   66.749\\
  32977 &   18.315 &    ... &   18.540 &   19.272 &   27.790 &   62.190 &   68.117\\
  41199 &   18.400 &    ... &   18.587 &   19.111 &   58.860 &   47.010 &   75.329\\
  31316 &   18.476 &    ... &   18.723 &   19.398 &  -33.530 &   67.550 &   75.414\\
  23960 &   18.565 &   18.200 &   18.757 &   19.209 &  -73.000 &  -22.760 &   76.466\\
  44882 &   18.006 &    ... &   18.278 &   19.397 &   92.550 &   13.970 &   93.598\\
\end{tabular} 
\end{center}
}


\clearpage

{
\hsize=15cm \vsize=22cm \hoffset -1cm

\begin{center}
\pagestyle{empty}

\small
{\bf Tab.4 The adopted global BSS sample} 
\vskip0.6truecm
\begin{tabular}{rlrrrrcccc}
\hline
 \multicolumn{1}{c}{N} &\multicolumn{1}{c}{id} &\multicolumn{1}{c}{V}
&\multicolumn{1}{c}{X} &\multicolumn{1}{c}{Y} 
&\multicolumn{1}{c}{r} &\multicolumn{1}{c}{GYBS} 
&\multicolumn{1}{c}{BHS} &\multicolumn{1}{c}{BPQ} &\multicolumn{1}{c}{F93}\\ 
\hline
   1& HST7821${\bf^\dagger}$ &   18.176 &   -1.040 &    0.610 &    1.206&1208&no&out&out\\
   2& HST5866${\bf^\dagger}$ &   17.863 &   -1.750 &    1.230 &    2.139&1141&no&out&out\\
   3& HST7570 &   18.685 &   -2.280 &   -1.060 &    2.514&1082&no&out&out\\
   4& HST5906${\bf^\dagger}$ &   17.510 &   -2.680 &   -0.190 &    2.687&1060&4131&out&out\\
   5& HST7785${\bf^\dagger}$ &   17.963 &   -1.110 &   -2.690 &    2.910&1196&no&out&out\\
   6& HST6674${\bf^\dagger}$ &   16.593 &    2.210 &    2.740 &    3.520&no&no&out&out\\
   7& HST7262 &   17.931 &   -0.400 &   -4.370 &    4.388&1259&no&out&out\\
   8& HST7613 &   17.581 &   -4.620 &   -0.680 &    4.670&no&no&out&out\\
   9& HST6430${\bf^\dagger}$ &   17.310 &   -1.930 &   -4.430 &    4.832&no&no&out&out\\
  10& HST5981${\bf^\dagger}$ &   18.016 &   -2.750 &   -5.260 &    5.935&no&no&no&out\\
  11& HST7678${\bf^\dagger}$ &   17.028 &   -6.060 &   -0.870 &    6.122&802&2304&out&out\\
  12& HST3736${\bf^\dagger}$ &   17.801 &    6.050 &    1.060 &    6.142&no&3841&out&out\\
  13& HST6832 &   18.509 &   -4.340 &   -4.520 &    6.266&no&no&no&out\\
  14& HST7339 &   18.072 &    6.820 &   -1.180 &    6.921&no&no&out&out\\
  15& HST6837 &   18.434 &   -5.230 &   -5.010 &    7.242&no&no&no&out\\
  16& HST6318${\bf^\dagger}$ &   18.166 &   -1.040 &    7.790 &    7.859&no&2906&out&out\\
  17& HST6910 &   18.509 &    7.920 &   -0.390 &    7.930&no&no&out&out\\
  18& HST7850${\bf^\dagger}$ &   18.308 &   -3.810 &   -7.120 &    8.075&966&no&80&out\\
  19& HST6979${\bf^\dagger}$ &   17.557 &    9.070 &   -1.470 &    9.188&2115&no&out&out\\
  20& HST7879 &   18.318 &    4.330 &   -8.280 &    9.344&1659&no&out&out\\
  21& HST7365 &   18.475 &   -2.190 &   -9.510 &    9.759&no&no&193&out\\
  22& HST6977 &   17.813 &   -7.900 &   -5.990 &    9.914&no&no&out&out\\
  23& HST5313${\bf^\dagger}$ &   17.637 &    9.670 &   -2.620 &   10.019&no&5662&out&out\\
  24&HST47941 &   18.225 &    8.760 &    4.970 &   10.072&no&no&out&out\\
  25& HST8009 &   17.866 &    5.340 &   -8.590 &   10.115&no&no&out&out\\
  26& HST6902 &   18.027 &   -2.350 &    9.900 &   10.175&no&no&out&out\\
  27& HST6352${\bf^\dagger}$ &   17.799 &    9.720 &   -4.800 &   10.841&2167&5061&out&out\\
  28& HST5229 &   18.376 &   -2.910 &   11.000 &   11.378&no&no&out&out\\
  29& HST7068 &   18.432 &    2.000 &  -11.370 &   11.545&no&no&361&out\\
  30&HST46237 &   17.860 &    8.170 &    8.190 &   11.568&no&no&out&out\\
  31& HST7069${\bf^\dagger}$ &   17.284 &    1.940 &  -11.490 &   11.653&1457&no&363&out\\
  32&HST47770 &   18.540 &    5.560 &   10.630 &   11.996&no&no&out&out\\
  33& HST2751 &   18.213 &   -2.770 &  -12.030 &   12.345&no&no&no&out\\
  34& HST5081 &   18.063 &    4.390 &  -12.040 &   12.815&no&no&no&out\\
  35& HST2926${\bf^\dagger}$ &   17.082 &   -8.150 &   -9.960 &   12.870&613&3487&out&out\\
  36& HST1769${\bf^\dagger}$  &   16.989 &   -0.940 &  -12.960 &   12.994&1202&2613&no&out\\
  37&HST50082 &   ... &   -9.230 &    9.280 &   13.089&no&no&out&out\\
  38&HST47750${\bf^\dagger}$  &   17.734 &   -1.030 &   13.280 &   13.320&no&2095&out&out\\
  39&HST47144 &   18.450 &   13.380 &    1.520 &   13.466&no&no&out&out\\
  40&HST47934${\bf^\dagger}$  &   18.114 &   10.070 &    9.090 &   13.566&no&6312&out&out\\
  41& HST2444${\bf^\dagger}$  &   17.655 &   -9.140 &  -10.120 &   13.636&557&no&out&out\\
  42& HST5681 &   18.188 &    3.170 &  -13.540 &   13.906&no&no&452&out\\
  43&HST51933 &   ... &  -11.810 &    7.990 &   14.259&no&no&out&out\\
\end{tabular} 
\end{center}
\newpage
\begin{center}
\pagestyle{empty}

\small
{\bf Tab.4 {\it (continued)}} 
\vskip0.6truecm
\begin{tabular}{rlrrrrcccc}
\hline
 \multicolumn{1}{c}{N} &\multicolumn{1}{c}{id} &\multicolumn{1}{c}{V}
&\multicolumn{1}{c}{X} &\multicolumn{1}{c}{Y} 
&\multicolumn{1}{c}{r} &\multicolumn{1}{c}{GYBS} 
&\multicolumn{1}{c}{BHS} &\multicolumn{1}{c}{BPQ} &\multicolumn{1}{c}{F93}\\ 
\hline
  44& HST1855${\bf^\dagger}$ &   18.133 &   -7.430 &  -12.430 &   14.481&673&no&44&out\\
  45& HST1292 &   18.275 &  -11.680 &   -8.570 &   14.487&no&no&out&out\\
  46& HST7321 &   18.285 &   13.520 &   -6.350 &   14.937&no&no&out&out\\
  47&HST48425${\bf^\dagger}$ &   18.294 &    9.640 &   11.420 &   14.945&no&no&out&out\\
  48& HST4925 &   18.522 &   11.040 &  -10.640 &   15.333&no&no&out&out\\
  49&HST51221 &   ... &  -14.700 &    4.550 &   15.388&no&no&out&out\\
  50&HST20075${\bf^\dagger}$ &   17.516 &  -12.860 &    9.880 &   16.217&no&no&out&out\\
  51&HST45876${\bf^\dagger}$ &   17.691 &   15.970 &    3.520 &   16.353&no&5404&out&out\\
  52& HST1244${\bf^\dagger}$ &   18.451 &   -0.120 &  -16.510 &   16.510&no&5444&392&out\\
  53&HST48269 &   17.614 &   16.530 &    0.380 &   16.534&no&no&out&out\\
  54&HST30161 &   18.600 &   -3.600 &   16.210 &   16.605&no&no&out&out\\
  55&HST20025 &   ... &   -8.020 &   14.750 &   16.789&no&no&out&out\\
  56& HST1398 &   18.017 &   -7.410 &  -15.280 &   16.982&no&no&out&out\\
  57& HST3407${\bf^\dagger}$ &   17.434 &  -11.740 &  -12.370 &   17.054&384&no&out&out\\
  58&HST48119${\bf^\dagger}$ &   17.657 &    8.490 &   15.030 &   17.262&no&no&out&out\\
  59& HST2922${\bf^\dagger}$ &   17.731 &  -17.310 &   -1.870 &   17.411&87&2070&no&out\\
  60& HST1284 &   18.162 &   -1.040 &  -17.420 &   17.451&no&no&out&out\\
  61&HST47970 &   18.284 &   17.530 &    0.070 &   17.530&no&no&out&out\\
  62&HST47973 &   18.222 &   17.580 &   -0.480 &   17.587&no&no&out&out\\
  63&HST48037 &   18.523 &   13.170 &   11.830 &   17.703&no&no&out&out\\
  64& HST3049 &   18.250 &  -17.520 &   -2.680 &   17.724&no&no&out&out\\
  65& HST7698${\bf^\dagger}$ &   17.426 &   11.840 &  -13.220 &   17.747&2323&4449&out&out\\
  66&HST43559 &   18.333 &    4.150 &   17.600 &   18.083&no&no&out&out\\
  67& HST2910 &   18.319 &    3.810 &  -17.880 &   18.281&1614&no&out&out\\
  68& HST1441${\bf^\dagger}$ &   17.918 &   -6.750 &  -17.110 &   18.393&722&no&no&out\\
  69&  HST519${\bf^\dagger}$ &   17.908 &  -16.200 &   -9.070 &   18.566&141&no&out&out\\
  70& HST2424 &   17.906 &    2.380 &  -18.420 &   18.573&1491&no&out&out\\
  71&HST20538 &   18.200 &  -12.520 &   13.730 &   18.581&no&no&out&out\\
  72& HST4361 &   18.280 &   10.420 &  -15.900 &   19.010&no&no&out&out\\
  73&HST47581 &   18.273 &    5.420 &   18.310 &   19.095&no&no&out&out\\
  74&HST47541 &   17.913 &   19.350 &    3.100 &   19.597&no&no&out&out\\
  75& HST1498${\bf^\dagger}$ &   16.918 &   -8.610 &  -17.960 &   19.917&576&1171&out&out\\
  76&HST44938 &   18.368 &    7.690 &   18.610 &   20.136&out&no&out&out\\
  77& HST4032${\bf^\dagger}$ &   17.857 &   15.340 &  -13.080 &   20.159&no&no&out&no\\
  78&HST47269${\bf^\dagger}$ &   18.005 &    3.760 &   20.010 &   20.360&out&5665&out&no\\
  79& HST5214${\bf^\dagger}$ &   17.867 &   19.200 &   -6.880 &   20.395&out&no&out&no\\
  80&HST20753 &   18.141 &  -18.250 &    9.290 &   20.478&out&no&out&no\\
  81&HST48349 &   17.955 &   20.640 &   -0.900 &   20.660&out&no&out&no\\
  82& HST7547 &   18.191 &   16.140 &  -13.090 &   20.781&out&no&out&no\\
  83&HST46285${\bf^\dagger}$ &   17.879 &   18.550 &    9.900 &   21.026&out&5078&out&no\\
  84&HST35380 &   18.310 &  -10.200 &   18.400 &   21.038&out&no&out&no\\
  85&HST20133 &   18.235 &  -21.070 &   -0.170 &   21.071&out&no&out&no\\
\end{tabular} 
\end{center}
\newpage

\begin{center}
\pagestyle{empty}

\small
{\bf Tab.4 {\it (continued)}} 
\vskip0.6truecm
\begin{tabular}{rlrrrrcccc}
\hline
 \multicolumn{1}{c}{N} &\multicolumn{1}{c}{id} &\multicolumn{1}{c}{V}
&\multicolumn{1}{c}{X} &\multicolumn{1}{c}{Y} 
&\multicolumn{1}{c}{r} &\multicolumn{1}{c}{GYBS} 
&\multicolumn{1}{c}{BHS} &\multicolumn{1}{c}{BPQ} &\multicolumn{1}{c}{F93}\\ 
\hline
  86&HST21131${\bf^\dagger}$ &   17.908 &  -17.040 &   13.700 &   21.864&out&no&out&no\\
  87&HST45855 &   18.204 &   19.420 &   10.660 &   22.153&out&no&out&no\\
  88& HST2979 &   18.270 &   -3.670 &  -22.590 &   22.886&out&no&out&no\\
  89&HST35187${\bf^\dagger}$ &   17.668 &   -8.090 &   21.600 &   23.065&out&1223&out&no\\
  90&HST35354 &   18.514 &   -4.160 &   22.770 &   23.147&out&2925&out&no\\
  91&HST21018${\bf^\dagger}$ &   18.014 &  -22.850 &    5.490 &   23.500&out&no&out&no\\
  92&HST20438 &   18.549 &  -24.150 &   -1.300 &   24.185&out&no&out&no\\
  93&HST47081${\bf^\dagger}$ &   17.998 &    8.760 &   23.160 &   24.761&out&1837&out&14\\
  94&HST48100 &   17.820 &    5.780 &   24.150 &   24.832&out&no&out&no\\
  95& HST3019 &   18.659 &  -19.130 &  -16.050 &   24.971&out&no&out&no\\
  96&HST48085 &   17.970 &   24.840 &    2.980 &   25.018&out&no&out&no\\
  97&HST48099 &   18.589 &    5.620 &   24.480 &   25.117&out&no&out&no\\
  98& HST3189${\bf^\dagger}$ &   17.887 &  -11.070 &  -23.040 &   25.561&out&no&out&no\\
  99&HST47164 &   18.476 &   23.510 &   11.410 &   26.133&out&no&out&no\\
 100& HST2070 &   18.546 &   -4.520 &  -25.900 &   26.291&out&no&out&no\\
 101&HST45066 &   17.812 &   18.520 &   19.000 &   26.533&out&no&out&no\\
 102& HST1076 &   18.456 &  -22.970 &  -13.440 &   26.613&out&no&out&no\\
 103& HST1534${\bf^\dagger}$ &   17.398 &    0.470 &  -27.420 &   27.424&out&no&out&no\\
 104&HST47590 &   18.262 &   19.550 &   19.270 &   27.451&out&no&out&no\\
 105&  F93-16${\bf^\dagger}$ &   17.833 &  -16.300 &  -22.400 &   27.703&121&2933&out&16\\
 106&HST35335${\bf^\dagger}$ &   17.864 &   -1.940 &   27.760 &   27.828&out&1400&out&no\\
 107&HST34816 &   18.048 &   -8.470 &   26.700 &   28.011&out&no&out&no\\
 108&HST47529 &   18.214 &   23.260 &   16.730 &   28.652&out&no&out&no\\
 109&HST45165 &   17.853 &   25.820 &   12.810 &   28.823&out&no&out&18\\
 110&HST48012 &   ... &   11.080 &   27.160 &   29.333&out&no&out&no\\
 111&HST41178 &   17.949 &   27.300 &   10.890 &   29.392&out&no&out&no\\
 112&HST48164${\bf^\dagger}$ &   17.437 &   28.250 &    9.000 &   29.649&out&no&out&no\\
 113&HST34446 &   18.776 &  -16.930 &   24.630 &   29.887&out&no&out&no\\
 114&HST34864 &   18.245 &   -0.330 &   30.150 &   30.152&out&no&out&no\\
 115&HST46418 &   17.956 &   30.290 &   -0.290 &   30.291&out&no&out&19\\
 116&HST30784${\bf^\dagger}$ &   18.167 &   -3.720 &   30.190 &   30.418&out&2405&out&no\\
 117&HST21270${\bf^\dagger}$ &   17.516 &  -30.470 &   -1.410 &   30.503&out&1559&out&20\\
 118&HST33562 &   18.186 &  -16.480 &   26.930 &   31.572&out&no&out&no\\
 119&HST44185 &   18.208 &   32.150 &    5.060 &   32.546&out&no&out&no\\
 120&HST47470${\bf^\dagger}$ &   17.609 &   32.350 &   -8.320 &   33.403&out&2322&out&no\\
 121&  F93-24${\bf^\dagger}$ &   17.043 &  -15.300 &  -30.000 &   33.676&out&851&out&24\\
 122&HST35310${\bf^\dagger}$ &   18.283 &    6.870 &   33.190 &   33.894&out&no&out&no\\
 123&HST47485 &   18.380 &   16.460 &   29.680 &   33.939&out&no&out&no\\
 124&HST46932 &   18.051 &   11.710 &   31.870 &   33.953&out&no&out&no\\
 125&HST33784${\bf^\dagger}$ &   17.555 &   -5.540 &   33.750 &   34.202&out&1204&out&25\\
 126&HST50012${\bf^\dagger}$ &   ... &  -30.240 &  -17.430 &   34.904&out&no&out&no\\
 127&  F93-26 &   18.545 &   26.900 &  -22.300 &   34.941&out&no&out&26\\
\end{tabular} 
\end{center}
\newpage

\begin{center}
\pagestyle{empty}

\small
{\bf Tab.4 {\it (continued)}} 
\vskip0.6truecm
\begin{tabular}{rlrrrrcccc}
\hline
 \multicolumn{1}{c}{N} &\multicolumn{1}{c}{id} &\multicolumn{1}{c}{V}
&\multicolumn{1}{c}{X} &\multicolumn{1}{c}{Y} 
&\multicolumn{1}{c}{r} &\multicolumn{1}{c}{GYBS} 
&\multicolumn{1}{c}{BHS} &\multicolumn{1}{c}{BPQ} &\multicolumn{1}{c}{F93}\\ 
\hline
 128&HST23174${\bf^\dagger}$ &   17.934 &  -30.510 &   18.460 &   35.660&out&2340&out&27\\
 129&HST47022${\bf^\dagger}$ &   17.045 &   36.470 &    2.750 &   36.574&out&no&out&29\\
 130&HST33861 &   18.184 &   -2.260 &   36.600 &   36.670&out&6282&out&28\\
 131&HST45828 &   18.311 &   28.260 &   24.840 &   37.625&out&no&out&no\\
 132&HST45007 &   18.447 &   23.660 &   29.860 &   38.097&out&1905&out&no\\
 133&HST23518${\bf^\dagger}$ &   18.195 &  -31.800 &   21.410 &   38.336&out&no&out&no\\
 134&HST45819${\bf^\dagger}$ &   17.305 &   26.630 &   27.660 &   38.396&out&no&out&no\\
 135&HST46434 &   18.695 &   39.330 &   -0.050 &   39.330&out&no&out&no\\
 136&  F93-32${\bf^\dagger}$ &   18.226 &  -29.000 &  -28.100 &   40.381&out&no&out&32\\
 137&HST21809 &   18.064 &  -40.500 &   -8.010 &   41.285&out&no&out&no\\
 138& BHS3236${\bf^\dagger}$ &   17.624 &  -20.810 &  -35.700 &   41.322&out&3236&out&no\\
 139&HST23822 &   17.897 &  -28.660 &   29.830 &   41.367&out&no&out&no\\
 140&  F93-34${\bf^\dagger}$ &   18.175 &   31.800 &  -26.900 &   41.652&out&2099&out&34\\
 141&  F93-36${\bf^\dagger}$ &   17.818 &   -3.600 &  -42.600 &   42.752&out&no&out&36\\
 142&HST23211${\bf^\dagger}$ &   17.078 &  -43.830 &    2.730 &   43.915&out&490&out&37\\
 143&HST22805 &   18.356 &  -43.880 &   -2.030 &   43.927&out&no&out&no\\
 144&HST46857 &   17.838 &   42.060 &   13.900 &   44.297&out&no&out&no\\
 145&HST45472 &   17.711 &   44.490 &   -4.020 &   44.671&out&no&out&no\\
 146&HST23181 &   17.917 &  -45.260 &    0.750 &   45.266&out&no&out&no\\
 147&HST23745${\bf^\dagger}$ &   17.838 &  -44.550 &    8.730 &   45.397&out&no&out&no\\
 148&HST44630${\bf^\dagger}$ &   18.461 &   40.320 &  -21.150 &   45.530&out&no&out&no\\
 149&HST34704${\bf^\dagger}$ &   17.675 &   13.930 &   43.360 &   45.543&out&no&out&no\\
 150&HST35241 &   18.212 &    9.500 &   45.370 &   46.354&out&no&out&no\\
 151&  F93-38${\bf^\dagger}$ &   18.235 &   38.200 &  -27.300 &   46.952&out&no&out&38\\
 152&HST31833 &   18.447 &    5.110 &   48.290 &   48.560&out&no&out&no\\
 153&  F93-39${\bf^\dagger}$ &   18.220 &    1.500 &  -48.800 &   48.823&out&no&out&39\\
 154&HST32122 &   18.426 &   17.710 &   45.610 &   48.928&out&no&out&no\\
 155&HST31956${\bf^\dagger}$ &   17.443 &    7.940 &   48.910 &   49.550&out&no&out&no\\
 156&HST24768${\bf^\dagger}$ &   16.628 &  -38.720 &   30.970 &   49.582&out&no&out&no\\
 157&HST42948 &   18.494 &   45.670 &  -22.980 &   51.126&out&no&out&no\\
 158&HST43517 &   18.220 &   25.920 &   46.460 &   53.201&out&no&out&no\\
 159&HST34564 &   17.980 &  -25.850 &   47.500 &   54.078&out&no&out&no\\
 160&HST34147${\bf^\dagger}$ &   18.107 &    3.410 &   54.180 &   54.287&out&no&out&no\\
 161&HST25134 &   18.219 &  -50.680 &   23.010 &   55.659&out&no&out&no\\
 162&HST32140${\bf^\dagger}$ &   17.340 &    2.950 &   58.400 &   58.474&out&235&out&42\\
 163&HST20674${\bf^\dagger}$ &   17.843 &  -50.870 &  -31.580 &   59.875&out&no&out&43\\
 164&HST25464${\bf^\dagger}$ &   18.326 &  -55.240 &   24.130 &   60.280&out&no&out&44\\
 165&  F93-45${\bf^\dagger}$ &   17.222 &  -10.700 &  -59.400 &   60.356&out&out&out&45\\
 166&HST24688${\bf^\dagger}$ &   18.202 &  -60.880 &    2.860 &   60.947&out&109&out&46\\
 167&  F93-48${\bf^\dagger}$ &   17.386 &    5.100 &  -61.700 &   61.910&out&out&out&48\\
 168&HST44881 &   18.179 &   56.040 &  -28.440 &   62.844&out&no&out&no\\
\end{tabular} 
\end{center}
\newpage

\begin{center}
\pagestyle{empty}

\small
{\bf Tab.4 {\it (continued)}} 
\vskip0.6truecm
\begin{tabular}{rlrrrrcccc}
\hline
 \multicolumn{1}{c}{N} &\multicolumn{1}{c}{id} &\multicolumn{1}{c}{V}
&\multicolumn{1}{c}{X} &\multicolumn{1}{c}{Y} 
&\multicolumn{1}{c}{r} &\multicolumn{1}{c}{GYBS} 
&\multicolumn{1}{c}{BHS} &\multicolumn{1}{c}{BPQ} &\multicolumn{1}{c}{F93}\\ 
\hline
 169&HST44584 &   18.257 &   65.090 &   13.220 &   66.419&out&no&out&no\\
 170&HST31898 &   18.183 &  -13.450 &   65.380 &   66.749&out&no&out&no\\
 171&HST32977 &   18.315 &   27.790 &   62.190 &   68.117&out&no&out&no\\
 172&HST21863${\bf^\dagger}$ &   18.075 &  -61.720 &  -33.870 &   70.403&out&2691&out&50\\
 173&HST23457${\bf^\dagger}$ &   17.870 &  -68.590 &  -24.110 &   72.704&out&out&out&51\\
 174&HST24338${\bf^\dagger}$ &   17.667 &  -71.860 &  -16.300 &   73.685&out&out&out&52\\
 175&  F93-53${\bf^\dagger}$ &   17.609 &   70.400 &  -22.500 &   73.908&out&out&out&53\\
 176&HST41199 &   18.400 &   58.860 &   47.010 &   75.329&out&out&out&no\\
 177&HST31316 &   18.476 &  -33.530 &   67.550 &   75.414&out&no&out&no\\
 178&  F93-54${\bf^\dagger}$ &   17.476 &  -62.700 &  -42.600 &   75.803&out&out&out&54\\
 179&HST26508${\bf^\dagger}$ &   17.824 &  -63.710 &   41.630 &   76.105&out&898&out&55\\
 180&  F93-56${\bf^\dagger}$ &   17.467 &   56.500 &  -51.200 &   76.248&out&out&out&56\\
 181&HST23960 &   18.565 &  -73.000 &  -22.760 &   76.466&out&out&out&no\\
 182&  F93-57${\bf^\dagger}$ &   17.960 &   34.000 &  -68.500 &   76.474&out&out&out&57\\
 183&HST35060${\bf^\dagger}$ &   16.856 &   -5.500 &   76.770 &   76.967&out&out&out&no\\
 184&HST26142${\bf^\dagger}$ &   17.288 &  -76.680 &   14.610 &   78.059&out&out&out&58\\
 185&  F93-59${\bf^\dagger}$ &   17.678 &   82.100 &  -22.600 &   85.154&out&out&out&59\\
 186&  F93-60${\bf^\dagger}$ &   17.404 &   86.700 &   -3.300 &   86.763&out&out&out&60\\
 187&  F93-61${\bf^\dagger}$ &   18.218 &   91.100 &   -8.400 &   91.486&out&out&out&61\\
 188&  F93-62${\bf^\dagger}$ &   18.173 &   79.300 &  -47.100 &   92.233&out&out&out&62\\
 189&  F93-63${\bf^\dagger}$ &   17.061 &   -9.700 &   92.700 &   93.206&out&out&out&63\\
 190&HST44882 &   18.006 &   92.550 &   13.970 &   93.598&out&out&out&no\\
 191&  F93-64 &   18.576 &   90.100 &  -27.400 &   94.174&out&out&out&out\\
 192&HST26972${\bf^\dagger}$ &   17.739 & -101.480 &   11.820 &  102.166&out&out&out&out\\
 193&  F93-65${\bf^\dagger}$ &   17.600 &  -99.000 &  -33.800 &  104.611&out&out&out&out\\
 194&  F93-66 &   18.409 &   90.200 &  -77.700 &  119.052&out&out&out&out\\
 195&  F93-67${\bf^\dagger}$ &   17.365 &   99.400 &  -96.400 &  138.468&out&out&out&out\\
 196&  F93-68${\bf^\dagger}$ &   18.112 &   29.600 &  167.900 &  170.489&out&out&out&out\\
 197&  F93-69${\bf^\dagger}$ &   18.277 & -188.500 &   65.700 &  199.621&out&out&out&out\\
 198&  F93-70${\bf^\dagger}$ &   17.477 &    8.000 & -201.000 &  201.159&out&out&out&out\\
 299&  F93-71 &   18.891 & -176.300 & -115.900 &  210.985&out&out&out&out\\
 200&  F93-72${\bf^\dagger}$ &   17.418 & -185.000 &  103.500 &  211.984&out&out&out&out\\
 201&  F93-74${\bf^\dagger}$ &   17.757 & -206.500 &   50.900 &  212.681&out&out&out&out\\
 202&  F93-73 &   18.724 &   70.000 &  201.000 &  212.840&out&out&out&out\\
 203&  F93-75 &   18.998 & -195.900 &   88.100 &  214.799&out&out&out&out\\
 204&  F93-76 &   18.723 & -207.700 &   68.300 &  218.642&out&out&out&out\\
 205&  F93-77 &   18.732 & -191.100 &  107.600 &  219.310&out&out&out&out\\
 206&  F93-78 &   18.838 &   62.000 &  213.000 &  221.840&out&out&out&out\\
 207&  F93-79 &   18.865 & -195.600 &  117.900 &  228.385&out&out&out&out\\
 208&  F93-80 &   18.540 &  207.000 &  100.000 &  229.889&out&out&out&out\\
 209&  F93-81${\bf^\dagger}$ &   17.611 &  112.000 &  202.000 &  230.972&out&out&out&out\\
 210&  F93-82${\bf^\dagger}$ &   18.124 &   65.000 & -224.000 &  233.240&out&out&out&out\\
\end{tabular} 
\end{center}
\newpage

\begin{center}
\pagestyle{empty}

\small
{\bf Tab.4 {\it (continued)}} 
\vskip0.6truecm
\begin{tabular}{rlrrrrcccc}
\hline
 \multicolumn{1}{c}{N} &\multicolumn{1}{c}{id} &\multicolumn{1}{c}{V}
&\multicolumn{1}{c}{X} &\multicolumn{1}{c}{Y} 
&\multicolumn{1}{c}{r} &\multicolumn{1}{c}{GYBS} 
&\multicolumn{1}{c}{BHS} &\multicolumn{1}{c}{BPQ} &\multicolumn{1}{c}{F93}\\ 
\hline
 211&  F93-83 &   18.572 &   66.000 & -225.000 &  234.480&out&out&out&out\\
 212&  F93-84 &   18.304 &  -82.000 &  222.000 &  236.660&out&out&out&out\\
 213&  F93-85${\bf^\dagger}$ &   18.033 &   37.000 & -237.000 &  239.871&out&out&out&out\\
 214&  F93-86${\bf^\dagger}$ &   17.912 & -205.600 & -126.900 &  241.609&out&out&out&out\\
 215&  F93-87 &   18.951 & -211.600 &  116.900 &  241.744&out&out&out&out\\
 216&  F93-88${\bf^\dagger}$ &   18.146 &   33.000 & -241.000 &  243.249&out&out&out&out\\
 217&  F93-89 &   18.914 &   61.000 &  239.000 &  246.662&out&out&out&out\\
 218&  F93-90 &   18.793 &  226.000 &  110.000 &  251.348&out&out&out&out\\
 219&  F93-91 &   18.942 &  200.000 &  155.000 &  253.032&out&out&out&out\\
 220&  F93-92 &   18.791 &  244.000 &   77.000 &  255.861&out&out&out&out\\
 221&  F93-93 &   19.040 &  -96.000 &  239.000 &  257.560&out&out&out&out\\
 222&  F93-94${\bf^\dagger}$ &   17.751 &  252.000 &  -63.000 &  259.756&out&out&out&out\\
 223&  F93-95 &   18.492 & -256.000 &   51.000 &  261.031&out&out&out&out\\
 224&  F93-96 &   18.640 & -154.000 & -211.000 &  261.222&out&out&out&out\\
 225&  F93-97 &   18.719 & -263.000 &   67.000 &  271.400&out&out&out&out\\
 226&  F93-98${\bf^\dagger}$ &   18.075 &   55.000 & -266.000 &  271.627&out&out&out&out\\
 227&  F93-99 &   18.390 &  -73.000 & -262.000 &  271.980&out&out&out&out\\
 228& F93-101 &   18.841 &  256.000 &  106.000 &  277.078&out&out&out&out\\
 229& F93-102 &   18.703 &   70.000 & -270.000 &  278.927&out&out&out&out\\
 230& F93-103 &   19.016 &  279.000 &   -1.000 &  279.002&out&out&out&out\\
 231& F93-104 &   18.426 &  257.000 &  111.000 &  279.946&out&out&out&out\\
 232& F93-100 &   19.016 &   92.000 & -265.000 &  280.516&out&out&out&out\\
 233& F93-105 &   18.404 & -164.000 & -235.000 &  286.568&out&out&out&out\\
 234& F93-106${\bf^\dagger}$ &   17.722 &   11.000 & -287.000 &  287.211&out&out&out&out\\
 235& F93-107 &   18.633 & -156.000 & -245.000 &  290.450&out&out&out&out\\
 236& F93-108${\bf^\dagger}$ &   18.200 & -286.000 &   52.000 &  290.689&out&out&out&out\\
 237& F93-109 &   18.633 &  104.000 & -272.000 &  291.204&out&out&out&out\\
 238& F93-110 &   18.390 & -114.000 &  278.000 &  300.466&out&out&out&out\\
 239& F93-111${\bf^\dagger}$ &   18.146 &   24.000 &  303.000 &  303.949&out&out&out&out\\
 240& F93-112${\bf^\dagger}$ &   18.200 & -281.000 & -118.000 &  304.770&out&out&out&out\\
 241& F93-113${\bf^\dagger}$ &   17.913 &  203.000 &  230.000 &  306.772&out&out&out&out\\
 242& F93-114 &   18.781 &  141.000 &  273.000 &  307.262&out&out&out&out\\
 243& F93-115${\bf^\dagger}$ &   17.524 & -189.000 &  244.000 &  308.637&out&out&out&out\\
 244& F93-116 &   18.408 &  312.000 &   34.000 &  313.847&out&out&out&out\\
 245& F93-117${\bf^\dagger}$ &   18.152 &  292.000 &  124.000 &  317.238&out&out&out&out\\
 246& F93-118${\bf^\dagger}$ &   17.823 & -231.000 & -225.000 &  322.469&out&out&out&out\\
 247& F93-119${\bf^\dagger}$ &   17.560 &  283.000 &  159.000 &  324.607&out&out&out&out\\
 248& F93-120 &   18.324 & -285.000 &  157.000 &  325.383&out&out&out&out\\
 249& F93-121 &   18.596 &  233.000 & -244.000 &  337.380&out&out&out&out\\
 250& F93-122 &   18.471 & -348.000 &   53.000 &  352.013&out&out&out&out\\
 251& S53-124 &   18.317 &    ...   &    ...   &  361.000&out&out&out&out\\
 252& S53-123${\bf^\dagger}$ &   18.477 &    ...   &    ...   &  361.000&out&out&out&out\\
 253& S53-125 &   19.057 &    ...   &    ...   &  370.000&out&out&out&out\\
\end{tabular} 
\end{center}
\newpage

\begin{center}
\pagestyle{empty}

\small
{\bf Tab.4 {\it (continued)}} 
\vskip0.6truecm
\begin{tabular}{rlrrrrcccc}
\hline
 \multicolumn{1}{c}{N} &\multicolumn{1}{c}{id} &\multicolumn{1}{c}{V}
&\multicolumn{1}{c}{X} &\multicolumn{1}{c}{Y} 
&\multicolumn{1}{c}{r} &\multicolumn{1}{c}{GYBS} 
&\multicolumn{1}{c}{BHS} &\multicolumn{1}{c}{BPQ} &\multicolumn{1}{c}{F93}\\ 
\hline
 254& F93-126${\bf^\dagger}$ &   17.791 &  122.000 &  351.000 &  371.598&out&out&out&out\\
 255& F93-127${\bf^\dagger}$ &   17.788 & -285.000 & -240.000 &  372.592&out&out&out&out\\
 256& F93-128${\bf^\dagger}$ &   17.380 &  156.000 &  350.000 &  383.192&out&out&out&out\\
 257& S53-129 &   18.707 &    ...   &    ...   &  410.000&out&out&out&out\\
 258& S53-130 &   18.717 &    ...   &    ...   &  410.000&out&out&out&out\\
 259& F93-131${\bf^\dagger}$ &   17.506 &  408.000 &  -52.000 &  411.300&out&out&out&out\\
 260& S53-132 &   18.417 &    ...   &    ...   &  420.000&out&out&out&out\\
 261& S53-133${\bf^\dagger}$ &   18.057 &    ...   &    ...   &  450.000&out&out&out&out\\
 262& S53-134${\bf^\dagger}$ &   18.117 &    ...   &    ...   &  460.000&out&out&out&out\\
 263& S53-135${\bf^\dagger}$ &   17.507 &    ...   &    ...   &  550.000&out&out&out&out\\
\end{tabular}
\end{center}
}

\end{document}